\documentclass[aps,nofootinbib,showpacs,preprintnumbers,amsmath,amssymb]{revtex4}
\usepackage{epsfig}

\begin{document}

\preprint{USM-TH-131, hep-ph/0210185, to appear in Phys.Rev.D}

\title{QCD effective action with dressing functions 
-- consistency checks in the perturbative regime}

\author{Gorazd Cveti\v{c}}
  \email{Gorazd.Cvetic@fis.utfsm.cl}
\author{Igor Kondrashuk}
 \email{Igor.Kondrashuk@fis.utfsm.cl}
\author{Iv\'an Schmidt}
 \email{Ivan.Schmidt@fis.utfsm.cl}
\affiliation{Dept.~of Physics, Universidad T\'ecnica
Federico Santa Mar\'{\i}a, Valpara\'{\i}so, Chile}

\date{\today}

\begin{abstract}
In a previous paper, we presented solution to the
Slavnov--Taylor identity for the QCD effective action, and 
argued that the action terms containing (anti)ghost fields
are unique. These terms have the same form as those in the 
classical action, but the gluon and (anti)ghost effective 
fields are convoluted with gluon and ghost dressing 
functions $G_A$ and $G_c$, the latter containing perturbative 
and nonperturbative effects (but not including the soliton-like
vacuum effects).
In the present work we show how the perturbative QCD (pQCD) 
can be incorporated into the framework of this action, 
and we present explicit one-loop pQCD expressions  
for $G_A$ and $G_c$.
We then go on to check the consistency of the obtained results 
by considering an antighost Dyson--Schwinger equation (DSE).
By solving the relations that result from the Legendre 
transformation leading to the effective action, we obtain 
the effective fields as power expansions of sources. We check 
explicitly that the aforementioned one-loop functions $G_A$ and 
$G_c$ fulfil the antighost DSE at the linear source level.
We further explicitly check that these one-loop $G_A$ and $G_c$
have the regularization-scale and momentum
dependence consistent with the antighost DSE at the
quadratic source level. These checks suggest that the 
the effective action with dressing functions 
represents a consistent framework for treating QCD,
at least at the one-loop level.
\end{abstract}

\maketitle

\section{Introduction}
\label{s1}

Nonperturbative methods are essential in order to rigorously
prove well accepted QCD phenomena such as confinement and
chiral symmetry breaking. In principle the effective
action contains all these effects, allowing for the 
calculation of all proper (one-particle-irreducible)
$n$-point Green functions. Furthermore, its minimum
gives the true quantum mechanical vacuum of the theory.
One way to obtain information about the effective
action is through the Slavnov--Taylor (ST) identity
\cite{ST,SF,Lee}, which is a consequence of the invariance of the
classical (tree level) action with respect to the BRST
transformation \cite{BRST,Becchi}.
The ST identity for the effective action
relates different functional
derivatives of this action with respect
to the effective fields. In the $N\!=\!1$
four-dimensional supersymmetric theory without matter, 
the form of the solution to the ST has been found
in Ref.~\cite{Kondrashuk:2000br}.  
In Ref.~\cite{KCS}, we presented a
solution to the ST identity for the
effective action of nonsupersymmetric non-Abelian
gauge theory. 

QCD is an example of nonsupersymmetric non-Abelian
gauge theory.
The solution has the form of the classical (QCD) action, 
but the fields entering it are convolutions of the
corresponding effective fields with pertaining
dressing functions. These functions
are supposed to contain the quantum effects of
the theory -- perturbative and nonperturbative,
with the possible exception of soliton-like vacuum effects.
Further, in that work we argued that the presented terms
of the solution which contain the ghost (and antighost)
fields are the only ones possible. 
Stated differently, the terms in
the effective action containing (anti)ghost terms
have the classical form and all the quantum effects
connected with Green functions with at least one
(anti)ghost leg must be contained in the ghost and
gluon dressing functions $G_A$ and $G_c$.

We also assume that those terms in the effective
action solution which contain no (anti)ghost fields
have essentially the classical form, but with the
fields convoluted with the dressing functions.
Further, the coupling parameter is running. 
In this case, we have not presented arguments
that such a solution to the ST identity for the
effective action is the only possible one.
The main attractive feature of this effective action
is that all the quantum effects are parametrized with
a {\em finite} number of dressing functions
and with the running function of the coupling.

A natural question appearing is whether the known
QCD results can be consistently incorporated into
the aforementioned effective action, i.e., into the
gluon, ghost and quark dressing functions. 
In this work we show how the known one-loop
perturbative QCD (pQCD) results are reflected
in this effective action. In fact, we find explicit
one-loop pQCD expressions for the ghost and gluon
dressing functions $G_A$ and $G_c$ by requiring
that the the correct known one-loop gluon and ghost
propagators be reproduced starting from our effective action.
It is not clear in advance whether our QCD effective
action framework with the obtained dressing
functions gives a self-consistent description
of QCD in the high-momentum (perturbative) regime.
We address this question of self-consistency
in the following way.
We first write down an antighost Dyson--Schwinger
equation (DSE) for QCD. Then we express the
effective action fields as a power series in the
(external) sources, 
by solving the relations that follow from the
Legendre transformation leading from the path integral
to the effective action. These power series
involve the aforementioned dressing functions.
We then insert these expansions in the
antighost DSE, with the one-loop dressing
functions $G_A$ and $G_c$. We explicitly
check that the regularization-scale and the momentum
dependence of these $G_A$ and $G_c$ are consistent
with the antighost DSE at the linear source
and the quadratic source level,\footnote{
The two DSE's involve indirectly three- and four-point
connected Green functions.} while at the linear source
level we even check the consistency of the
constant (``finite'') parts of the one-loop $G_A$ and $G_c$.
All this strongly
suggests that our effective action framework
is consistent with pQCD in the corresponding
(high) momentum region. It further keeps open the
realistic possibility that the dressing functions
of our effective action represent a sufficient 
framework for including not just all the
(high momentum) perturbative, but also
(lower momentum) nonperturbative quantum effects.

In Sec.~\ref{sec:dse} we recapitulate the results of 
Ref.~\cite{KCS} and present the aforementioned antighost DSE.
In Sec.~\ref{sec:dress} we deduce the one-loop
gluon and ghost dressing functions $G_A$ and $G_c$
from requiring the correct one-loop gluon
and ghost propagators.
In Sec.~\ref{sec:Legendre} we write down the
relations arising from the Legendre transformation
leading to the effective action, and obtain the effective
fields as power expansions in sources.
In Sec.~\ref{sec:dsecheck} we explicitly check that
the obtained one-loop dressing functions $G_A$ and $G_c$ 
are consistent with the antighost DSE at the linear level
of sources. At the quadratic level of sources, we explicitly
check that the regularization-scale and momentum dependence
of the one-loop $G_A$ and $G_c$ is consistent with
the antighost DSE. In Sec.~\ref{sec:summary} we
summarize our results.  

\section{The effective action and the antighost Dyson--Schwinger
equation}
\label{sec:dse}

In order to fix notations, we first write the well-known
classical action of QCD, in the Lorentz gauge
\begin{eqnarray}
 S_{\rm {QCD}}[A,b,c,q,\bar{q}] =
\lefteqn{ 
\int~d^{4}x~ \Big\{ 
- \frac{1}{2 g^2}~{\rm Tr} 
\left[F_{\mu \nu}(A(x))F^{\mu \nu}(A(x)) \right]
 + \bar{q}_{(j)} ~i~\gamma^{\mu} \nabla_{\mu}(A) q_{(j)} }
\nonumber\\
& & - \frac{1}{\alpha}~{\rm Tr} \left( 
\left[\partial_{\mu} A^{\mu}(x) \right]^2 \right)
   - 2~{\rm Tr} \left[~i~b(x)\partial^{\mu}~\nabla_{\mu}(A)~c(x)\right] 
\Big\} \ .
\label{SQCD}
\end{eqnarray}
Here, $A$, $b$, $c$, and $q_j$ are the gluon, antighost,
ghost, and quark (j'th flavor) fields, and $\alpha$ is the
gauge parameter. For simplicity we omit the quark masses. 
For all the fields $X$ 
in the adjoint representation ($X = A, b, c$) we use the notation
$X = X^a T^a$, where $T^a = \lambda^a/2$ are the gauge group
generators ($N_c \times N_c$ hermitian matrices in the color space)
with the normalization
\begin{subequations}
\label{Tnormall}
\begin{eqnarray}
{\rm Tr} \left( T^a T^b \right) &=& 
\frac{1}{2} \delta^{ab} \ , \quad 
(T^a)^{\dagger} = T^a \ , 
\label{Tnorm1}
\\
\left[ T^b,T^c\right] &=& i f^{abc} T^a\ .
\label{Tnorm2}
\end{eqnarray}
\end{subequations}
The antighost and ghost fields are hermitian Grassmann
numbers $b^{\dagger} = b$, $c^{\dagger} = c$. 
The covariant derivatives in the ghost part in (\ref{SQCD}) are 
\begin{subequations}
\begin{eqnarray}
\nabla_{\mu}(A) c & = & 
\partial_{\mu} c + i [A_{\mu}, c]  \ ,
\label{Dc}
\\
F_{\mu \nu} & = & \partial_{\mu} A_{\nu} - \partial_{\nu} A_{\mu}
+ i [A_{\mu}, A_{\nu} ] \ ,
\label{FA}
\\
\nabla_{\mu}(A) q_{(j)} & = & \partial_{\mu} q_{(j)} + 
i A_{\mu} q_{(j)} \ .
\label{Dq}
\end{eqnarray}
\end{subequations}
The fields $X^a$, $q_{(j)}$,
the coupling $g$, and the gauge-fixing parameter $\alpha$
in the action (\ref{SQCD}) are bare, i.e.,
the theory has a large (but finite) UV momentum cutoff $\Lambda$
\begin{equation}
X \equiv X^{(\Lambda)} \ , \quad 
q_{(j)} \equiv q_{(j)}^{(\Lambda)} \ , \quad
g \equiv g(\Lambda) \ , \quad 
\alpha \equiv \alpha^{(\Lambda)} \equiv g^2(\Lambda) \xi^{(\Lambda)} \ ,
\label{bare}
\end{equation}
and the cutoff is implemented through a gauge invariant
regularization, e.g.~dimensional regularization.

A solution to the Slavnov--Taylor (ST) identity for the
effective action $\Gamma$ has a similar form 
as the classical action (\ref{SQCD}). However, the
fields are modified $X \mapsto {\tilde X}$
\cite{KCS}.~\footnote{
The exception is the gauge-fixing part which is
unchanged under quantum corrections.} 
In particular, the terms containing the (anti)ghost fields and
the quark fields have just the same form as
in the classical action, but with $X \mapsto {\tilde X}$,
and the pure gauge boson terms have gauge invariant form,
but with $A \mapsto {\tilde A}$
\begin{eqnarray}
\Gamma_{\rm {QCD}}[A,b,c,q,\bar{q}] =
\lefteqn{ 
\int~d^{4}x~ \Big\{ 
{\cal L}_1^{\rm eff} \left( {\tilde A}(x) \right) + 
{\cal L}_2^{\rm eff} \left( {\tilde q}_{(j)}(x), 
{\bar {\tilde q}}_{(j)}(x),{\tilde A}(x) \right)
}
\nonumber\\
& & - \frac{1}{\alpha}~{\rm Tr} \left( 
\left[\partial_{\mu} A^{\mu}(x) \right]^2 \right)
   - 2~{\rm Tr} \left[~i~{\tilde b}(x)~
\partial^{\mu}\nabla_{\mu}({\tilde A})~{\tilde c}(x)\right] 
\Big\} \ .
\label{GQCD}
\end{eqnarray}
Here, ${\cal L}_1^{\rm eff}({\tilde A}(x))$ is a gauge invariant
combination of the gluon field ${\tilde A}$.
The modified fields ${\tilde X}$ appearing in (\ref{GQCD})
are the original fields convoluted with the corresponding 
dressing functions $G_X$
\begin{subequations}
\label{tildes}
\begin{eqnarray}
{\tilde A}_{\mu}^a(x) &=& \left( G_A^{-1} \circ A_{\mu}^a \right) (x)
= \int~d^{4}x^{\prime}~G_A^{-1}(x - x^{\prime})
A_{\mu}^a(x^{\prime}) \ ,
\label{tildeA} \\
{\tilde c}^a(x) &=& \left( G_c^{-1} \circ c^a \right) (x)
= \int~d^{4}x^{\prime}~G_c^{-1}(x - x^{\prime})
c^a(x^{\prime}) \ ,
\label{tildec} \\
{\tilde q}_{(j) \xi}(x) & = & 
\left( G_q^{-1} \circ q_{(j)} \right)_{\xi} (x)
= \int~d^{4}x^{\prime}~
G_q^{-1}(x - x^{\prime})_{\xi \xi^{\prime}} 
q_{(j) \xi^{\prime}}(x^{\prime}) \ ,
\label{tildeq} \\
{\tilde b}^a(x) &=& \left( G_A \circ b^a \right) (x)
= \int~d^{4}x^{\prime}~G_A(x - x^{\prime})
b^a(x^{\prime}) \ .
\label{tildeb} 
\end{eqnarray}
\end{subequations}
Notice that the dressing functions $G_X$ appearing in relations 
(\ref{tildeA}) and (\ref{tildeb}) are the same ($G_A$).
In (\ref{tildeq}), the spinor structure is reflected in
indices $\xi, \xi^{\prime} = 1,2,3,4$. The spacetime structure
of the dressing functions is
$\langle x | G_X | x^{\prime} \rangle = G_X(x - x^{\prime})$ 
due to translational invariance. These functions are
real for $X\!=\!A,c$, and are in general complex 
$4 \times 4$ spinor matrices for $X\!=\!q_{(j)}$.
Furthermore, the Fourier transforms of the dressing functions
$G_A$ and $G_c$
\begin{equation}
G_X(k^2) = \int~d^{4}x~\exp(i k \cdot x)~G_X(x)
\label{FTGX}
\end{equation}
are functions of $k^2$ due to Lorentz invariance. As a
consequence, $G_X(x) = G_X(-x)$ ($X=A,c$).
Apart from the aforementioned invariance properites of the
dressing functions, the ST-identity for the effective action
does not tell us anything about these functions.

We have argued in Ref.~\cite{KCS} that the terms containing
the (anti)ghost fields in the solution (\ref{GQCD}) of the
ST identity have a unique form as given in Eq.~({\ref{GQCD}).
This was a consequence of the assumption
of Ref.~\cite{KCS} that the term $\sim\!L c^2$ of the
effective action fulfils separately the ST identity
for the effective action. Here, $L$ is an auxiliary
(nonpropagating) background field which couples
at the tree level to the BRST transformation of the
ghost field $c$ in the form $2 {\rm Tr}[L(x) c(x)^2]$ in
the (classical) Lagrangian density. 
The aforementioned assumption
was then shown to imply the following structure 
of the $\sim\!L c^2$ term of the effective action:
\begin{equation}
\int d^{4}x ~2 ~{\rm Tr} \left[ 
\left( G_c \circ L \right)(x) 
\left( G_c^{-1} \circ c \right)(x)
\left( G_c^{-1} \circ c \right)(x) \right] \ ,
\label{Lc2}
\end{equation}
where $G_c(x)$ is, a priori, an arbitrary dressing function.
An argument was given in Ref.~\cite{KCS} in favor of the 
aforementioned assumption. 
Furthermore, arguments were presented there 
which suggest the structure (\ref{Lc2}) as the only
consistent $\sim L c^2$ structure even when the mentioned 
assumption is not adopted.
The structure of this correlator was then shown in
Ref.~\cite{KCS} to restrict the gluon-ghost-antighost term 
to the one written in Eq.~(\ref{GQCD}),
and the physical part of the effective action was 
shown to be gauge-invariant in terms of the effective fields 
convoluted with the dressing functions (\ref{tildes}) 
${\tilde X} = G^{-1}_X \circ X$.

The gauge-fixing term in (\ref{GQCD}) is
also unique, due to its known insensitivity to quantum corrections.
In (\ref{GQCD}), the expression ${\cal L}_2^{\rm eff}$ which
contains the quark fields is gauge-invariant combination
of ${\tilde q}_{(j)}$, ${\bar {\tilde q}}_{(j)}$ and ${\tilde A}$. 
In our further considerations in this paper, the precise
form of these terms will not matter as we will
concentrate on the question of consistency of the (anti)ghost
and gluon sectors of the effective action (\ref{GQCD})
in the perturbative region.

The pure gluon term ${\cal L}_1^{\rm eff}({\tilde A}(x))$ 
in the effective action has a form somewhat different
from the form of the pure gluon term
in the classical action (\ref{SQCD}). This
should be expected because the latter term is the
only one which contains, in addition to the field(s),
the coupling parameter $g$. This parameter is running.
Thus, it is natural to expect that the latter parameter 
in the effective action is not just normalized to a specific 
value $g(\mu_0)$, where $\mu_0$ is a specific, in principle arbitrary,
renormalization scale. Rather, when going into the
Fourier-transformed (momentum) space, we should associate
to each $k$-mode of the dressed gluon field ${\tilde A}$
the value of $g(k)$. Specifically, while the pure
gluon sector in the classical action (\ref{SQCD}) has the
following form in the momentum space:
\begin{equation}
S_{\rm gl}[A] = - \frac{1}{2 g^2}~
\int~\frac{d^{4}k}{(2 \pi)^4}~{\rm Tr}
\left[ F_{\mu \nu}(A)(k)~F^{\mu \nu}(A)(-k) \right] \ ,
\label{Sgl}
\end{equation} 
where $F_{\mu \nu}(A)(k)$ denotes the Fourier
transform of $F_{\mu \nu}(A)(x) \equiv F_{\mu \nu}(A(x))$,
the pure gluon sector in the effective action (\ref{GQCD})
contains the analogous term
\begin{equation}
\Gamma_{\rm gl,1}[A] =  - \frac{1}{2}~
\int~\frac{d^{4}k}{(2 \pi)^4}~\frac{1}{g^2(-k^2)}~
{\rm Tr} \left[ 
F_{\mu \nu}({\tilde A})(k)~F^{\mu \nu}({\tilde A})(-k) 
\right] \ .
\label{Ggl1}
\end{equation}  
However, in contrast to (\ref{Sgl}), the expression
(\ref{Ggl1}) is not gauge invariant when 
${\tilde A} \mapsto A$, although the full
gluon sector of the effective action (\ref{GQCD})
must have this property as a consequence of the
ST-identity. Thus, gluon contributions additional
to those in (\ref{Ggl1}) must appear. 
In order to illustrate how to obtain these additional
gluon terms, we will find them explicitly 
in the one-loop perturbative approximation.
In this case, we have the relation
\begin{subequations}
\label{1lRGEall}
\begin{eqnarray}
\frac{1}{g^2(-k^2)} &=& \frac{1}{g^2(\mu_0^2)} +
\frac{\beta_0}{4 \pi^2} \ln \left(
\frac{- k^2}{\mu_0^2} \right) \ ,
\label{1lRGErun}
\\
{\rm where} \  \beta_0 &=& \frac{1}{12} ( 11 N_c - 2 n_f) \ .
\label{beta0}
\end{eqnarray}
\end{subequations}
Here, $\beta_0$ is the one-loop coefficient for
the beta-function for $\alpha_s\!\equiv\!g^2/(4 \pi)$,
$N_c\!=\!3$ is the number of colors, $n_f$ is the
number of active quark flavors. Using the one-loop
evolution form (\ref{1lRGErun}) in (\ref{Ggl1}),
we can rewrite the latter in the form of the
following spacetime integrals:
\begin{eqnarray}
\Gamma_{\rm gl,1}^{\rm 1-loop}[A] &=& 
- \frac{1}{2 g^2(\mu_0)}~\int~d^{4}x~{\rm Tr} 
\left[F_{\mu \nu}({\tilde A})(x)~F^{\mu \nu}({\tilde A})(x) \right]
\nonumber\\
&& - \frac{1}{2} \frac{\beta_0}{4 \pi^2}~\int~d^{4}x~{\rm Tr} 
\left[ F_{\mu \nu}({\tilde A})(x)~\ln 
\left( \frac{\partial^2 }{\mu_0^2} \right)~
F^{\mu \nu}({\tilde A})(x) \right] \ .
\label{Ggl11l}
\end{eqnarray}
The integrand in the second term is not gauge invariant
when ${\tilde A} \mapsto A$. However, the minimal extension
of the latter integrand to recover the aforementioned
property is to replace 
$\partial_{\mu} \mapsto \nabla_{\mu}({\tilde A})$
\begin{eqnarray}
\Gamma_{\rm gl}^{\rm 1-loop}[A] &=& 
- \frac{1}{2 g^2(\mu_0)}~\int~d^{4}x~{\rm Tr} 
\left[F_{\mu \nu}({\tilde A})(x)~F^{\mu \nu}({\tilde A})(x) \right]
\nonumber\\
&&- \frac{1}{2} \frac{\beta_0}{4 \pi^2}~\int~d^{4}x~{\rm Tr} 
\left[ F_{\mu \nu}({\tilde A})(x)~\ln 
\left( \frac{ \nabla^2({\tilde A}(x)) }{\mu_0^2} \right)~
F^{\mu \nu}({\tilde A})(x) \right] \ .
\label{Ggl1l}
\end{eqnarray}
By expanding the new logarithm in (\ref{Ggl1l})
around its value at $A\!=\!0$, we can immediately see
that the new terms, i.e., the difference
of the integrands of the second integrals in
(\ref{Ggl11l}) [$\leftrightarrow$ (\ref{Ggl1})]
and (\ref{Ggl1l}), are terms at least cubic in gluon
fields and of the type $1/\partial$,
i.e., in momentum space these are power-suppressed
terms of the type $1/k$. Such terms will not be
relevant in our perturbative considerations, only the 
terms that depend logarithmically on the momenta will be.
Therefore, in the perturbative considerations,
we can consider the form (\ref{Ggl1}) to
represent the entire gluonic sector of the
effective action. 

In Appendix \ref{app:lndel} we show explicitly how 
Eq.~(\ref{Ggl11l}) follows from Eqs.~(\ref{Ggl1})
and (\ref{1lRGEall}), present clear expressions
for the logarithmic operators appearing in
the integrals in Eqs.~(\ref{Ggl11l}) and (\ref{Ggl1l}),
and show (tilde-)gauge invariance of the expression
(\ref{Ggl1l}). Further, in that Appendix we argue
how to proceed in the general (nonperturbative) case, i.e., 
when $1/g(-k^2)$ is a general function
of $k^2$, in order to obtain from (\ref{Ggl1}) the
(tilde-)gauge invariant version of the gluonic
effective action.

For further use in the manuscript, we will write here
a general Dyson--Schwinger-type equation (DSE) which
the QCD path integral has to fulfill. The QCD path integral 
$Z \equiv \exp(- i W)$ is defined as
\begin{subequations}
\label{ZQCD2}
\begin{eqnarray}
\exp \left( - i W[J,\rho,\eta,{\bar j},j] \right) &=&
\int[d A] [d b] [d c] [d q_{(j)}] [d {\bar q}_{(j)}]~
\exp\left( i S_{\rm {QCD}}[A,b,c,q,\bar{q}] + 
i \Xi[A,b,c,q,\bar{q};J,\rho,\eta,{\bar j},j] \right) ,
\label{ZQCD}
\end{eqnarray}
with
\begin{eqnarray}
\Xi[A,b,c,q,\bar{q};J,\rho,\eta,{\bar j},j] & = & 
2~{\rm Tr}~\left( \int~d^{4}x~J_{\mu} A^{\mu}
+ i~\int~d^{4}x~\rho(x) b(x) 
+ i~\int~d^{4}x~\eta(x) c(x) \right)
\nonumber\\
&& + \int~d^{4}x~({\bar q}_{(j)} j_{(j)} + {\bar j}_{(j)} q_{(j)} ) \ ,
\label{jpart}
\end{eqnarray}
\end{subequations}
where $J,\rho,\eta,j,{\bar j}$ are the (external) sources
for the corresponding fields. The path integral $Z$ (\ref{ZQCD})
is just a function of sources, with the fields integrated out.
Therefore, in particular, $Z$ is invariant under the 
(infinitesimal) shift of the antighost fields $b(x)
\mapsto b(x) + \epsilon(x)$.
This invariance can be rewritten,
after some straightforward algebraic manipulations,\footnote{
Care should be taken about the Grassmannian anticommuting
character of $\rho$ and $b$.}
in the form
\begin{eqnarray}
\partial^2 c^a(x) - f^{abc}~\partial^{\mu}
\left[ A_{\mu}^b(x) c^c(x) + 
\frac{ \delta^2 W}{\delta J^{\mu}_{b}(x)~\delta \eta^c(x)} \right]
& = & - \rho^a(x) \ .
\label{bDSE0}
\end{eqnarray}
Moreover, due to the relation 
$\delta W/\delta \eta^c(x) = - i c^c(x)$ which follows from
the Legendre transformation that connects $W$ and $\Gamma_{\rm QCD}$,
this relation
can be rewritten as
\begin{eqnarray}
\partial^2 c^a(x) - f^{abc}~\partial^{\mu}
\left[ A_{\mu}^b(x) c^c(x) - i~ 
\frac{ \delta c^c(x)}{\delta J^{\mu}_{b}(x)} \right]
& = & - \rho^a(x) \ .
\label{bDSE}
\end{eqnarray}
Here it is understood that the effective fields
$c$ and $A$ are functions of sources.
We will call the identity (\ref{bDSE}) the antighost Dyson--Schwinger
equation ($b$-DSE). The usual $b$-DSE equation is obtained  
by applying to (\ref{bDSE}) the variation
$\delta/(\delta \rho^d(y))$
and then setting all the sources and fields equal to zero.
In (\ref{bDSE}), the fields and the sources are in
general nonzero.

\section{PQCD predictions for the dressing functions}
\label{sec:dress}

We will here deduce some implications of
perturbative QCD (pQCD) for the hitherto unknown
dressing functions appearing in (\ref{tildes}).
We will write explicit expressions for them
as predicted by one-loop pQCD.

The pQCD prediction for the full gluon propagator is
\begin{eqnarray}
 \left( \frac{\delta^2 W}
{\delta J^{a_1 \mu_1} \delta J^{a_2 \mu_2} }\right) (k) & = &
\left( \frac{\delta^2 \Gamma}
{\delta A_{\mu_1}^{a_1} \delta A_{\mu_2}^{a_2}}\right)^{-1}\!\! (k)
= g^2(\Lambda)~Z_3(\Lambda,K) ~D^{a_1 a_2 (K)}_{\mu_1 \mu_2 }(k) \ ,
\label{gpropQCD}
\end{eqnarray}
where $Z_3(\Lambda,K)$ appears as a factor in 
the gluon field rescaling function
for the theory with cutoff $K$ ($K^2\!=\!- k^2$)
\begin{equation}
A_{\mu}^{(\Lambda)} = Z_g(\Lambda,K)~Z_3^{1/2}(\Lambda,K)~
A_{\mu}^{(K)} \ ,
\label{Aresc}
\end{equation}
and $Z_g(\Lambda,K)$ is the rescaling (``running'') function
for the QCD coupling parameter
\begin{equation}
g(\Lambda) = Z_g(\Lambda,K)~g(K) \ .
\label{gresc}
\end{equation} 
Here we use a simplified notation for the running $g$:
$g(K) \equiv g(-k^2) $, $g(\Lambda) \equiv g(\Lambda^2)$.
The rescaling functions $Z_3$ and $Z_g$ are calculable in pQCD. 
In (\ref{gpropQCD}),
$D_{\mu_1 \mu_2}^{a_1 a_2 (K)}(k)$ is the usual gluon propagator
in the theory with cutoff $K$ ($K^2 = - k^2$), i.e.,
the usual tree level propagator
\begin{equation}
D^{a_1 a_2 (K)}_{\mu_1 \mu_2}(k) = \delta_{a_1 a_2}
\frac{(-1)}{k^2} \left( g_{\mu_1 \mu_2}
- \frac{ k_{\mu_1} k_{\mu_2} }{k^2} + \xi^{(K)} 
\frac{ k_{\mu_1} k_{\mu_2} }{k^2}
\right) \ ,
\label{gproptree}
\end{equation}
and $\xi^{(K)} \equiv \alpha^{(K)}/g^2(K)$ ($\sim g^0)$
is the usual gauge parameter. On the other hand, the
gluon propagator as predicted by the effective action
(\ref{GQCD}), with the gluon part (\ref{Ggl1}), is 
\begin{equation}
\left( \frac{\delta^2 \Gamma}
{\delta A_{\mu_1}^{a_1} \delta A_{\mu_2}^{a_2}}\right)^{-1} (k)
= \delta_{a_1 a_2}  \frac{(-1)}{k^2} 
\left[ g^2(K)~\left(G_A(k^2)\right)^2 \left(
g_{\mu_1 \mu_2} - \frac{ k_{\mu_1} k_{\mu_2} }{k^2} \right) 
+ \alpha \frac{ k_{\mu_1} k_{\mu_2} }{k^2}
\right] \ .
\label{gpropG}
\end{equation}
We stress here that this is the full gluon propagator
in our framework. The additional nonperturbative terms in 
the gluonic sector of the effective action
which appeared in Eq.~(\ref{Ggl1l})
as a consequence of the substitution $\partial \mapsto
\nabla({\tilde A}(x))$ in the logarithm, are terms
at least cubic in the gluon fields ($\sim\!(A(k))^3/k$)
and thus do not affect the gluon propagator which is
determined solely by the $\sim\!(A(k))^2$ terms of the
gluon part (\ref{Ggl1}).
Comparing the transversal parts of
the pQCD propagator (\ref{gpropQCD}) 
and of our propagator (\ref{gpropG}),
we conclude
\begin{eqnarray}
G_A(k^2) &=& \left( \frac{g(\Lambda)}{g(K)} \right) 
~Z^{1/2}_3(\Lambda,K)  =  Z_g(\Lambda,K)~Z^{1/2}_3(\Lambda,K) \ . 
\label{GApQCD}
\end{eqnarray}
The longitudinal parts of the
pQCD propagator (\ref{gpropQCD}) 
and of our propagator (\ref{gpropG})
agree now automatically, because of
the relation (\ref{bare}) for the
gauge parameter $\alpha$ and the
known relation
\begin{equation}
\xi^{(\Lambda)} = Z_3(\Lambda,K) ~\xi^{(K)} \ .
\label{xirun}
\end{equation}

Now we analogously consider the ghost propagator.
The pQCD prediction for the full ghost propagator is 
\begin{eqnarray}
 \left( \frac{\delta^2 W}
{\delta \eta^{a_1} \delta \rho^{a_2} } \right) (k)  & = &
\left( \frac{\delta^2 \Gamma}
{\delta b^{a_2} \delta c^{a_1}}\right)^{-1}\!\!(k)
= \delta_{a_1 a_2} Z_2^{(c)}(\Lambda,K) \frac{i}{k^2} \ ,
\label{cpropQCD}
\end{eqnarray}
where $Z_2^{(c)}(\Lambda,K)$ is the ghost and antighost
field rescaling function for the theory with cutoff
$K$ ($K^2\!=\!-k^2$)
\begin{equation}
c^{(\Lambda)} = Z_2^{(c) 1/2}(\Lambda,K) ~c^{(K)} \ , \quad
b^{(\Lambda)} = Z_2^{(c) 1/2}(\Lambda,K) ~b^{(K)} \ .
\label{bcresc}
\end{equation}
On the other hand, our effective action (\ref{GQCD})
predicts
\begin{equation}
\left( \frac{\delta^2 \Gamma}
{\delta b^{a_2} \delta c^{a_1}}\right)^{-1}\!\!(k) =
\delta_{a_1 a_2} \left( \frac{G_c(k^2)}{G_A(k^2)} \right)~\frac{i}{k^2} \ .
\label{cpropG}
\end{equation}
Comparing this with the pQCD result (\ref{cpropQCD}) gives us
\begin{eqnarray}
G_c(k^2) &=& Z_2^{(c)}(\Lambda,K)~G_A(k^2) 
\nonumber\\
&=& Z_2^{(c)}(\Lambda,K)~Z_3^{1/2}(\Lambda,K)~Z_g(\Lambda,K) \ .
\label{GcpQCD}
\end{eqnarray}
Although we will not use the explicit form of the quark dressing 
function (matrix) $G_q$, we mention here that
completely analogous considerations involving the
quark propagator give us a possible (but not unique)
pQCD form for $G_q$
\begin{equation}
G_q(k^2)_{\xi \xi^{\prime}} = 
Z^{1/2}_2(\Lambda,K) ~\exp\left( i \theta(k^2) \right)
\delta_{\xi \xi^{\prime}} \ ,
\label{GqpQCD}
\end{equation}
where $\theta(k^2)$ is a momentum-dependent phase,
and $Z_2(\Lambda,K)$ is the quark field rescaling
function ($K^2\!=\!-k^2$)
\begin{equation}
q_{(j)}^{(\Lambda)} = Z_2^{1/2}(\Lambda,K) ~q_{(j)}^{(K)} \ .
\label{qresc}
\end{equation}
The rescaling (``running'') functions $Z_3, Z_2^{(c)}, Z_2$
and $Z_g$ are all computable in pQCD, and at the
one-loop level are given by
\begin{subequations}
\label{Z1lall}
\begin{eqnarray}
Z_3(\Lambda,K) & = & 1 + \frac{1}{\pi} \alpha_s
\left[ \kappa_3 \ln \left( \frac{ \Lambda^2 }{K^2} \right) +
{\tilde \kappa}_3 \right] + {\cal O}(\alpha_s^2) \ ,
\label{Z31l}
\\
Z_2^{(c)}(\Lambda,K)& =& 1 + \frac{1}{\pi} \alpha_s 
\left[ \kappa_2^{(c)} \ln \left( \frac{ \Lambda^2 }{K^2} \right) 
+ {\tilde \kappa}_2^{(c)} \right]
+ {\cal O}(\alpha_s^2) \ ,
\label{Z2c1l}
\\
Z_2(\Lambda,K) & = & 1 + \frac{1}{\pi} \alpha_s
\left[ \kappa_2 \ln \left( \frac{ \Lambda^2 }{K^2} \right) 
+ {\tilde \kappa}_2 \right]
+ {\cal O}(\alpha_s^2) \ ,
\label{Z21l}
\\
Z_g(\Lambda,K) & = & 1 + \frac{1}{\pi} \alpha_s 
\left[ \kappa_g \ln \left( \frac{ \Lambda^2 }{K^2} \right) 
+ {\tilde \kappa}_g \right]
+ {\cal O}(\alpha_s^2) \ ,
\label{Zg1l}
\end{eqnarray}
\end{subequations}
where $\alpha_s \equiv g^2/(4 \pi)$ and
the coefficients $\kappa_X$ are well known
\begin{subequations}
\label{kall}
\begin{eqnarray}
\kappa_3 & = &  \frac{1}{24} \left[ (13 - 3 \xi) N_c - 4 n_f \right] \ ,
\label{k3}
\\
\kappa_2^{(c)}&=& \frac{1}{16} (3 - \xi) N_c \ ,
\label{k3c}
\\
\kappa_2 &=& - \xi \frac{1}{4} \frac{(N_c^2 -1)}{2 N_c} \ ,
\label{k2}
\\
\kappa_g &=& - \frac{1}{2} \beta_0 = - \frac{1}{24} (11 N_c - 2 n_f) \ .
\label{kg}
\end{eqnarray}
\end{subequations}
Here, $N_c\!=\!3$ is the number of colors, $n_f$ is the number
of active quark flavors, and $\beta_0$ is from Eq.~(\ref{beta0}).
Further, the ``finite'' one-loop parts, which are $\Lambda$-
and $k$-independent, are known as well. For example, in
the ${\overline {\rm MS}}$ renormalization scheme, using
dimensional regularization, we have
\begin{subequations}
\label{tkall}
\begin{eqnarray}
{\tilde \kappa}_3 & = & \frac{1}{4} \left\{ \left[
\frac{31}{9} - \left( 1 - \xi \right) + 
\frac{1}{4} \left( 1 - \xi \right)^2 \right] N_c
- \frac{10}{9} n_f \right\} \ ,
\label{tk3}
\\
{\tilde \kappa}_2^{(c)} & = & \frac{1}{4} N_c \ .
\label{tk2c}
\end{eqnarray}
\end{subequations}
The constants ${\tilde \kappa}_2$ and ${\tilde \kappa}_g$
are also known, but we will not need them at the level of
our analysis. The ``finite'' part constant (\ref{tk3})
is written for the case when the active quarks are massless.
The two constants (\ref{tkall}) can be
obtained by evaluating the dimensionally regularized
gluon and ghost self-energy parts -- see, for example, 
the book \cite{Muta} [Eqs.~(A.21) and (A.23) there].
In the above formulas, we denoted by $\Lambda$
the ${\overline {\rm MS}}$ UV cutoff which is obtained by the
formal substitution $[ 2/(4\!-\!D) - \gamma + \ln ( 4 \pi) ]
\mapsto \ln( \Lambda^2)$, where $D$ $(\to 4)$
is the dimension in the regularization. The coupling $\alpha_s$
is, in principle, the ``bare'' ${\overline {\rm MS}}$
coupling $\alpha_s = \alpha_s(\Lambda^2;{\overline {\rm MS}})$
here. However,  the momentum dependence of $\alpha_s$ 
in expressions (\ref{Z1lall}) is irrelevant at the $\sim\!\alpha_s$ level
as it affects only $\sim\!\alpha_s^2$ terms.

Inserting the known one-loop results (\ref{Z1lall})--(\ref{tkall})
into (\ref{GApQCD}) and (\ref{GcpQCD}), we obtain the
one-loop pQCD predictions for the gluon and ghost
dressing functions $G_A$ and $G_c$
\begin{subequations}
\label{GX1lall}
\begin{eqnarray}
G_X(k^2) & = & \left[ 1 + \frac{1}{2 \pi} \alpha_s
\left[ \kappa_X \ln \left( \frac{ \Lambda^2 }{-k^2} \right) 
+ {\tilde \kappa}_X \right]
+ {\cal O}(\alpha_s^2) 
\right] 
\ , \quad (X=A,c) \ ,
\label{GX1l}
\\
\kappa_A &=& (\kappa_3 - \beta_0) = - \frac{1}{8} (3 + \xi) N_c \ ,
\label{kA}
\\
\kappa_c &=& (2 \kappa_2^{(c)} + \kappa_3 - \beta_0) = 
- \frac{1}{4} \xi N_c \ ,
\label{kc}
\\
{\tilde \kappa}_A &=& {\tilde \kappa}_3 + 2 {\tilde \kappa}_g \ ,
\quad
\label{tkA}
\\
{\tilde \kappa}_c -  {\tilde \kappa}_A & = & 2 {\tilde {\kappa}}_2^{(c)}
= \frac{1}{2} N_c \ . 
\label{tkcmA}
\end{eqnarray}
\end{subequations}

The freedom of choice of the UV regularization scale $\Lambda$
is reflected in the dressing functions $G_X$ which depend on 
$\Lambda$
\begin{equation}
G_X(k^2) = G_X^{(\Lambda)}(k^2) \ .
\label{GXLmu0}
\end{equation}

Up until now, we considered the renormalization scale in
the dressing functions to be a very high UV cutoff scale
$\Lambda \gg \Lambda_{\rm QCD}$. But the
arbitrariness of $\Lambda$ implies that, in the Wilsonian sense,
we could have started from the classical action with a lower
cutoff $\mu < \Lambda$, where $\mu$ is a renormalization
scale. Then, in all our formulas,
we would have to make the simple substitution
$\Lambda \mapsto \mu$.
The resulting rescaling of the dressing functions
under the change $\Lambda \mapsto \mu$ is then
\begin{subequations}
\label{Grescall}
\begin{eqnarray}
G_A^{(\Lambda)}& = &Z_g(\Lambda,\mu) ~Z_3^{1/2}(\Lambda,\mu)~
G_A^{(\mu)} \ ,
\label{GAresc}
\\
G_c^{(\Lambda)}& = &Z_2^{(c)}(\Lambda,\mu)~Z_3^{1/2}(\Lambda,\mu)~
Z_g(\Lambda,\mu)~G_c^{(\mu)} \ .
\label{Gcresc}
\end{eqnarray}
\end{subequations}
It is then straightforward to see that the
field ${\tilde A}$ and the combination ${\tilde b} {\tilde c}$,
appearing in the effective action (\ref{GQCD}), are independent
of the choice of the renormalization scale $\mu$.
This can be seen via Eqs.~(\ref{Grescall}),
(\ref{Aresc}) and (\ref{bcresc}), in the latter two replacing
$K \mapsto \mu$. Analogously, it can be shown that
the combination ${\bar {\tilde q}}_{(j)} {\tilde q}_{(j)}$
appearing in the effective action is independent of $\mu$.

\section{Fields as power expansions in sources}
\label{sec:Legendre}

In this Section we will write down the 
transformations associated with the Legendre transformation
which relates the path integral function $W$ (\ref{ZQCD})
with the effective action $\Gamma_{\rm QCD}$ (\ref{GQCD}). 
We will then solve them, thus obtaining the
effective fields $c$ and $A$ as power expansions in sources.
We recall that the following Legendre transformation
must relate $W$ with $\Gamma_{\rm QCD}$:
\begin{eqnarray}
\Gamma[A,b,c,q,{\bar q}] & = & - W[J,\rho,\eta,{\bar j},j] -
\Xi[A,b,c,q,\bar{q};J,\rho,\eta,{\bar j},j] \ ,
\label{Legendre}
\end{eqnarray}
where $\Xi$ is given in Eq.~(\ref{jpart}).
{}From here follows the $b$(antighost)-relation
\begin{equation}
\frac{\delta \Gamma}{\delta b^a(x)}
\left( \equiv \frac{\delta {\tilde b}^b}{\delta b^a} \circ 
\frac{\delta \Gamma}{\delta {\tilde b}^b} (x) \right)
= i \rho^a(x) \ .
\label{bL1}
\end{equation}
By Eqs.~(\ref{tildeb}) and (\ref{GQCD}) we have
\begin{equation} 
\frac{\delta {\tilde b}^b(y)}{\delta b^a(x)} = \delta_{ab}
G_A(y-x) \ , \quad
\frac{\delta \Gamma}{\delta {\tilde b}^b(y)} =
-i~\left( \partial^{\mu} \nabla_{\mu}({\tilde A})~{\tilde c}(y)
\right)^b \ .
\label{bL2}
\end{equation} 
Applying to (\ref{bL1}) the convolution
$i~G_A^{-1} \circ$ from the left, and using the aforementioned
symmetry $G_A^{-1}(-x) = G_A^{-1}(x)$, we then obtain
\begin{equation}
\partial^2 {\tilde c}^a(x) - f^{abc} \partial^{\mu}
\left( {\tilde A}^b_{\mu}(x) {\tilde c}^c(x) \right) =
- \int~d^{4}x^{\prime} G^{-1}_A(x - x^{\prime})
\rho^a(x^{\prime}) \ .
\label{bLx}
\end{equation}
This is the Legendre-related $b$-relation in
spacetime coordinates.
Applying the Fourier transformation of type (\ref{FTGX})
to this relation, we obtain the following 
Legendre-related $b$(antighost)-relation
in four-momentum coordinates, involving the
corresponding Fourier transformed functions:
\begin{equation}
- k^2~\frac{c^a(k)}{G_c(k^2)} + i~f^{abc} k^{\mu}~
\int~\frac{d^{4} k_1}{(2 \pi)^4}
\frac{A^b_{\mu}(k_1)}{G_A(k_1^2)}
\frac{c^c(k-k_1)}{G_c((k-k_1)^2)} =
- \frac{\rho^a(k)}{G_A(k^2)} \ .
\label{bLk}
\end{equation}

Completely analogously, we obtain the following
Legendre-related $c$(ghost)-relation 
in four-momentum space:
\begin{equation}
-k^2~G_A(k^2)~b^a(k) + i~f^{abc}
\int~\frac{d^{4} k_1}{(2 \pi)^4}
\frac{G_A((k-k_1)^2)}{G_A(k_1^2)} (k - k_1)^{\mu}
A^b_{\mu}(k_1)b^c(k-k_1) = G_c(k^2)~\eta^a(k) \ .
\label{cLk}
\end{equation}

Very analogous, but algebraically more involved
manipulations of the Legendre-related $A$(gluon)-relation
\begin{equation}
\frac{\delta \Gamma}{\delta A^{a_1}_{\mu_1}(x)}
\left( \equiv 
\frac{\delta {\tilde A}^{a_2}_{\mu_2}}
{\delta A^{a_1}_{\mu_1}} \circ 
\frac{\delta \Gamma}{\delta {\tilde A}^{a_2}_{\mu_2} } (x) \right)
= - J_{a_1}^{\mu_1}(x) 
\label{AL1}
\end{equation}
lead to the following explicit form of the Legendre-related
$A$(gluon)-relation in four-momentum coordinates:
\begin{eqnarray}
&& - \frac{1}{g^2(K) G_A(k^2)} \Bigg\{
k^2 \left[ \left(
g^{\mu_1 \mu_2} - \frac{k^{\mu_1} k^{\mu_2}}{k^2}
\right)  
+ \frac{g^2(K)}{\alpha} \left( G_A(k^2) \right)^2
\frac{k^{\mu_1} k^{\mu_2}}{k^2}
\right]
A^{a_1}_{\mu_2}(k) 
\nonumber\\
& & + i~ f^{a_1 a_2 a_3} G_A(k^2)~\int~\frac{d^{4} k_1}{(2 \pi)^4}
\frac{1}{G_A(k_1^2) G_A((k-k_1)^2)} 
\nonumber\\
&& \times \left[ \left( 
k_1^{\mu_1} A^{a_3 \mu_2}(k_1) -
k_1^{\mu_2} A^{a_3 \mu_1}(k_1) \right) A^{a_2}_{\mu_2}(k-k_1)
- k^{\mu_2} A^{a_2}_{\mu_2}(k_1) A^{a_3 \mu_1}(k-k_1)
\right]
\nonumber\\
&& + f^{a_1 a_2 b} f^{a_3 a_4 b} 
G_A(k^2)~\int~\frac{d^{4} k_1 d^{4} k_2}{(2 \pi)^8}
\frac{1}{G_A(k_1^2) G_A(k_2^2) G_A((k-k_1-k_2)^2)}
A^{a_2 \mu_2}(k_1) A^{a_3 \mu_1}(k_2) A^{a_4}_{\mu_2}(k-k_1-k_2)
\Bigg\}
\nonumber\\
&& + f^{a_1 a_2 a_3} \int~\frac{d^{4} k_1}{(2 \pi)^4}
\frac{G_A(k_1^2)}{G_c((k-k_1)^2)} k_1^{\mu_1} 
b^{a_2}(k_1) c^{a_3}(k-k_1) 
\nonumber\\
&& + ~\int~\frac{d^{4} k_1}{(2 \pi)^4}~
\left( {\bar q}_{(j)}(k_1) \gamma^0 G^{-1}_{q_1}(k) \gamma^0 \right)
\gamma^{\mu_1} T^{a_1}
\left( G^{-1}_q(k-k_1) q(k-k_1) \right) =
- ~G_a(k^2) J_{a_1}^{\mu_1}(k) \ .
\label{ALk}
\end{eqnarray}
The ${\bar q}$-relation from the Legendre transformation would 
involve, analogously, terms $\sim q_{(j)}$ and 
$\sim A q_{(j)}$ on the left-hand side
and the quark current $j_{(j)}$ on the right-hand side.
In the above relations, and in the rest of the article,
we omit for simplicity any notational reference to the
dependence of the effective fields, the
dressing functions, and the gauge parameter $\alpha$
on the UV cutoff $\Lambda$.

The Legendre-related relations (\ref{bLk}), (\ref{cLk}), 
and (\ref{ALk})
connect the effective fields with the sources.
They allow us to obtain the effective
fields as expansions in powers of sources
\begin{equation}
X = X^{(1)} + X^{(2)} + X^{(3)} + \cdots
\label{Xexp}
\end{equation}
Here we denote by $X^{(n)}$ the part
of the field $X$ of power $n$ in the sources
($J_a^{\mu}, \rho, \eta, {\bar j}_{(j)}, j_{(j)}$).
We assume that the fields go to zero when the
sources go to zero ($X^{(0)} = 0$), i.e.,
we assume that there are no soliton-like vacuum effects.
When inserting these expansions in the Legendre-related 
relations, and requiring that each coefficient
in the expansion fulfill the equation, we
obtain an infinite series of Legendre-related equations 
for $X^{(n)}$'s which can be solved successively.
For example, the $b$-relation (\ref{bLk})
thus gives the series of equations
\begin{subequations}
\label{bLkall}
\begin{eqnarray}
&&- \frac{k^2}{G_c(k^2)} c^a(k)^{(1)} = 
- \frac{1}{G_A(k^2)} {\rho}^a(k) \ ,
\label{bLk1}
\\
&&- \frac{k^2}{G_c(k^2)} c^a(k)^{(2)} + i~f^{abc} k^{\mu}~
\int~\frac{d^{4} k_1}{(2 \pi)^4}
\frac{1}{G_A(k_1^2) G_c((k-k_1)^2)}
A^b_{\mu}(k_1)^{(1)} c^c(k-k_1)^{(1)} = 0 \ ,
\label{bLk2}
\\
&&- \frac{k^2}{G_c(k^2)} c^a(k)^{(3)} + i~f^{abc} k^{\mu}~
\int~\frac{d^{4} k_1}{(2 \pi)^4}
\nonumber\\
&&\;\;\;\;\;\;\;\;\;\;\;\;\;\;\;\;
\times
\frac{1}{G_A(k_1^2) G_c((k - k_1)^2)}
\left[ A^b_{\mu}(k_1)^{(1)} c^c(k\!-\!k_1)^{(2)} + 
A^b_{\mu}(k_1)^{(2)} c^c(k-k_1)^{(1)} \right] = 0 \ ,
\label{bLk3}
\end{eqnarray}
\end{subequations}
etc. Solving these equations successively for
$c^{(1)}, c^{(2)}, c^{(3)}, \ldots$, we obtain
\begin{subequations}
\label{call}
\begin{eqnarray}
&& c^a(k)^{(1)} = \frac{G_c(k^2)}{G_A(k^2)} 
\frac{1}{k^2} {\rho}^a(k) \ ,
\label{c1}
\\
&& c^a(k)^{(2)} = i~f^{abc} G_c(k^2) \frac{k^{\mu}}{k^2}~
\int~\frac{d^{4} k_1}{(2 \pi)^4}
\frac{1}{G_A(k_1^2) G_c((k-k_1)^2) (k-k_1)^2}
A^b_{\mu}(k_1)^{(1)} {\rho}^c(k-k_1) \ ,
\label{c2}
\\
&& c^a(k)^{(3)} = i~f^{abc} G_c(k^2) \frac{k^{\mu}}{k^2}~
\int~\frac{d^{4} k_1}{(2 \pi)^4}
\nonumber\\
&&\;\;\;\;\;\;\;\;\;\;\;\;\;\;\;\;
\times
\frac{1}{G_A(k_1^2) G_c((k-k_1)^2)}
\left[ A^b_{\mu}(k_1)^{(1)} c^c(k\!-\!k_1)^{(2)} + 
A^b_{\mu}(k_1)^{(2)} c^c(k-k_1)^{(1)} \right] \ ,
\label{c3}
\end{eqnarray}
\end{subequations}
etc.
In an analogous, though algebraically more involved way, 
by inserting expansions
(\ref{Xexp}) into the Legendre-related $A$-relation (\ref{ALk})
we obtain expressions for $A^{(n)}$ 
\begin{subequations}
\label{Aall}
\begin{eqnarray}
A^{a_1}_{\mu_1}(k)^{(1)} & = &
g^2(K) \left( G_A(k^2) \right)^2 \frac{1}{k^2}
P_{\mu_1 \mu_2}(k) J_{a_1}^{\mu_2}(k) \ ,
\label{A1}
\\
A^{a_1}_{\mu_1}(k)^{(2)} & = &
i~f^{a_1 a_2 a_3} g^4(K) \frac{G_A(k^2)}{k^2}
P_{\mu_1 \mu_2}(k)~\int~\frac{d^{4} k_1}{(2 \pi)^4}
\frac{G_A(k_1^2) G_A((k-k_1)^2)}{k_1^2 (k-k_1)^2}
\nonumber\\
&& \!\!\!\!\!\!\!\!\!\!\!\!\!\!
\times
\left[ \left( k_1^{\mu_2} P_{\mu_3 \mu_4}(k_1) -
(k_1+k)_{\mu_3} P^{\mu_2}_{\;\;\;\mu_4}(k_1) \right) 
P^{\mu_3}_{\;\;\;\mu_5}(k-k_1) \right]
J_{a_2}^{\mu_4}(k_1) J_{a_3}^{\mu_5}(k-k_1) + 
{\cal O}(\eta \rho, {\bar j}_{(j)} j_{(j)})
\ ,
\label{A2}
\end{eqnarray}
\end{subequations}
etc., where we denoted by $P_{\mu \nu}$ the function proportional
to the gluon propagator (\ref{gpropG})
\begin{eqnarray}
P_{\mu_1 \mu_2}(k) & \equiv &
\left[ \left( g_{\mu_1 \mu_2} - \frac{k_{\mu_1} k_{\mu_2}}{k^2}
\right) +\frac{\alpha}{g^2(K)} \frac{1}{(G_A(k^2))^2}
\frac{k_{\mu_1} k_{\mu_2}}{k^2} \right] 
= \left[ \left( g_{\mu_1 \mu_2} - \frac{k_{\mu_1} k_{\mu_2}}{k^2}
\right) +\xi^{(K)} \frac{k_{\mu_1} k_{\mu_2}}{k^2} \right] 
\ .
\label{Pdef}
\end{eqnarray}
In the analysis of the next Section, we will not need
the explicit terms ${\cal O}(\eta \rho, {\bar j}_{(j)} j_{(j)})$
in (\ref{A2}). They originate from terms $\sim c^{(1)} b^{(1)}$
and ${\bar q}_{(j)}^{(1)} q_{(j)}^{(1)}$.
We can obtain the coefficient functions
$b^{(n)}$, $q^{(n)}$, ${\bar q}^{(n)}$ in a similar way, 
but we will not need them in the following analysis.
The cubic-in-sources part $c^{(3)}$ of $c$
can be obtained by inserting in the
expression (\ref{c3}) the explicit solutions
({\ref{c1})--(\ref{c2}) and ({\ref{A1})--(\ref{A2}).

\section{Checking the antighost DS equation}
\label{sec:dsecheck}

In this Section we will perform a consistency check of
the effective action framework (\ref{GQCD}) for the
high momenta (pQCD) region. In Section \ref{sec:dress}
we deduced the dressing functions for the region of
high momenta by requiring that they, in the
framework of (\ref{GQCD}), should give us the
known pQCD two-point Green functions.
In particular, the known one-loop pQCD expressions 
for the gluon and ghost propagators resulted
in the high-momentum behavior
(\ref{GX1lall}) for the
dressing functions $G_A$ and $G_c$.
On the other hand, in the expansion in sources
various Dyson--Schwinger equations (DSE's) can give
us (infinite) series of equations
for the dressing functions. For example, the
antighost DSE ($b$-DSE) (\ref{bDSE})
gives, after inserting into it for the effective
fields the power expansions in sources discussed
in the previous Section, one equation for each power
of sources
\begin{subequations}
\label{bDSxall}
\begin{eqnarray}
&&\partial^2 c^a(x)^{(1)} + i~f^{abc} \partial^{\mu}_{(x)}
\left(
\frac{\delta c^c(x)^{(2)}}{\delta J^{\mu}_{b}(x)} 
\right) = - {\rho}^a(x) \ ,
\label{bDS1x}
\\
&&\partial^2 c^a(x)^{(2)} - f^{abc} \partial^{\mu}_{(x)}
\left( A^{b}_{\mu}(x)^{(1)} c^c(x)^{(1)} \right)
+ i~f^{abc}\partial^{\mu}_{(x)} \left( 
\frac{\delta c^c(x)^{(3)}}{\delta J^{\mu}_{b}(x)} 
\right) = 0 \ ,
\label{bDS2x} 
\end{eqnarray}
\end{subequations}
etc. In part A of this Section, we will check explicitly
that the aforementioned one-loop pQCD solutions
(\ref{GX1lall}) fulfil Eq.~(\ref{bDS1x}). In part B
we will check explicitly that their dependence
on the UV cutoff $\Lambda$ and on the momentum $k$
is compatible with Eq.~(\ref{bDS2x}).

We first apply the Fourier transformations to these
relations, obtaining relations for the corresponding
Fourier-transformed functions
\begin{subequations}
\label{bDSall}
\begin{eqnarray}
&& - k^2 c^a(k)^{(1)} + f^{abc} k^{\mu}
\left( \frac{\delta c^{c (2)}}{\delta J^{\mu}_{b}} \right) (k) =
- {\rho}^a(k) \ ,
\label{bDS1}
\\
&&- k^2 c^a(k)^{(2)} + i~f^{abc} k^{\mu}
~\int~\frac{d^{4} k_1}{(2 \pi)^4} A^{b}_{\mu}(k_1)^{(1)}
c^c(k-k_1)^{(1)} + f^{abc} k^{\mu}
\left( \frac{\delta c^{c (3)}}{\delta J^{\mu}_b} \right) (k) = 0 \ ,
\label{bDS2}
\end{eqnarray}
\end{subequations}
etc.
The variational derivatives of the type
$(\delta c^{(n)}/\delta J)(k)$ in Eqs.~(\ref{bDSall})
denote simply the Fourier transforms, with respect to $x$,
of the functions $(\delta c(x)^{(n)}/\delta J(x))$.

We note that the term $\delta c^{(2)}/\delta J$
in the linear-in-sources DSE (\ref{bDS1x})
contains implicitly the three-point connected
Green function $\delta^2 c^{(2)}/(\delta \rho \delta J)
= i~ \delta^3 W/(\delta \rho \delta \eta \delta J)
\sim \langle b c A \rangle$. Analogously, the
term $\delta c^{(3)}/\delta J$ in the
quadratic-in-sources DSE (\ref{bDS2x})
contains implicitly the four-point connected
Green function $\delta^4 W/(\delta \rho \delta \eta \delta J \delta J)
\sim \langle b c A A \rangle$.

\subsection{Linear-in-sources DS equation}
\label{subsec:lcDS}

First we will show, in our effective action framework
(\ref{GQCD}) and (\ref{Ggl1}), that 
the one-loop dressing functions (\ref{GX1lall}) are
consistent with the linear-in-sources DS equation (\ref{bDS1}).
When we use in Eq.~(\ref{bDS1}) the expressions
(\ref{c1}), (\ref{c2}), and (\ref{A1}), the common
factor $\rho^a(k)$ factorizes out and we end up
with the equivalent relation (we multiply with $-G_A(k^2)$)
\begin{eqnarray}
&& G_c(k^2) + i~g^2(K) N_c \frac{1}{k^2}
~\int^{(\Lambda)}~\frac{d^{4} k_1}{(2 \pi)^4}
~\frac{G_c(k_1^2) G_A((k-k_1)^2)}{k_1^2 (k-k_1)^2}
k^{\mu_1} P_{\mu_1 \mu_2}(k-k_1) k_1^{\mu_2} = G_A(k^2) \ .
\label{bDS12}
\end{eqnarray}
All the squared momenta in the denominators,
here and in the rest of the article,
are understood to have $+i\epsilon$
added to ensure causality.
The integral must be understood as having been regularized
with a gauge-invariant cutoff $\Lambda$, which can
be implemented with dimensional regularization
(see Sec.~\ref{sec:dress} and Appendix \ref{app:regint}).

Further, we insert in (\ref{bDS12}) the
one-loop results (\ref{GX1lall}),
and thus are allowed to keep only terms up to
order ${\cal O}(\alpha_s^1)$. This means that
we replace in the integral $ G_c(k_1^2) G_A((k-k_1)^2) \mapsto 1$
and $\xi^{(K-K_1)} \mapsto \xi^{(\Lambda)}$ and
obtain the equivalent (one-loop) relation
\begin{eqnarray}
G_A(k^2)\!-\!G_c(k^2) & = &
i~4 \pi \alpha_s(K) N_c \frac{1}{k^2} 
~\int^{(\Lambda)}~\frac{d^{4} k_1}{(2 \pi)^4}
~\frac{1}{k_1^2 (k\!-\!k_1)^2}
k^{\mu_1} 
\left[ g_{\mu_1 \mu_2} + (\xi^{(\Lambda)}\!-\!1)
\frac{(k\!-\!k_1)_{\mu_1} (k\!-\!k_1)_{\mu_2}}{(k-k_1)^2}
\right] k_1^{\mu_2} .
\label{bDS13}
\end{eqnarray}
In Appendix \ref{app:regint}, we explicitly evaluated the 
integral on the right-hand side of DSE (\ref{bDS13}),
with the result given in Eq. (\ref{bDS13rhs}). On the other hand,
the one-loop expression results (\ref{GX1lall})
for the dressing functions give for the left-hand side
of DSE (\ref{bDS13})
\begin{equation}
G_A(k^2)\!-\!G_c(k^2) = 
- \alpha_s(K^2) N_c \frac{1}{16 \pi} \left[ (3\!-\!\xi^{(\Lambda)} )
\ln \left( \frac{\Lambda^2}{-k^2} \right) + 4 \right]
+ {\cal O}(\alpha_s^2)
\ .
\label{bDS13lhs}
\end{equation}
This is just the same as the result (\ref{bDS13rhs}) for
the right-hand side of DSE (\ref{bDS13}), at the 
one-loop level at which we are working.
This shows explicitly, in our effective action
framework, that the one--loop dressing functions (\ref{GX1lall})
are consistent with the DSE (\ref{bDS12}), or equivalently with
the first (linear in sources) DSE (\ref{bDS1})
in the series of the antighost DSE's 
(\ref{bDS1}), (\ref{bDS2}), etc.

\subsection{Quadratic-in-sources DS equation}
\label{subsec:qcDS}

We now proceed to show, in our effective action
framework (\ref{GQCD}) and (\ref{Ggl1}), that the
$\Lambda^2$- and $k^2$-dependent parts
of the one-loop dressing functions $G_X$ (\ref{GX1lall})
are consistent with
the quadratic-in-sources antighost DSE (\ref{bDS2}).
The explicit form of the first two terms in this DSE,
in terms of the dressing functions and sources, can be 
obtained by straightforward insertion of the expressions
(\ref{c1}), (\ref{c2}) and (\ref{A1})
\begin{eqnarray}
&&- k^2 ~c^a(k)^{(2)} + i~f^{abc} ~k^{\mu}
~\int^{(\Lambda)}~\frac{d^{4} k_1}{(2 \pi)^4} 
A^{b}_{\mu}(k_1)^{(1)} c^c(k-k_1)^{(1)} =
\nonumber\\
&& - i~g^2(K)~k_{\mu}~\int^{(\Lambda)}~\frac{d^{4} k_1}{(2 \pi)^4}
\frac{1}{k_1^2 (k - k_1)^2} \left[
\frac{G_c(k^2)}{G_c( (k-k_1)^2) G_A(k_1^2)} - 1 \right]
\left( f^{abc} J^{\mu}_b(k_1) {\rho}^c(k-k_1) \right)
\ .
\label{bDS21}
\end{eqnarray}
We recall that in our convention, $A \sim g$, $J \sim g^{-1}$,
${\rho} \sim g^0$. But the expression in brackets in the 
integral is $\sim g^2$, thus the expression (\ref{bDS21}) is
$\sim g^3 \sim \alpha_s^{3/2}$. Using the one-loop expressions 
(\ref{GX1lall}) for the dressing functions, 
the expression in brackets is
\begin{equation}
\left[
\frac{G_c(k^2)}{G_c( (k-k_1)^2) G_A(k_1^2)} - 1 \right] =
- \frac{\kappa_A}{2 \pi} \alpha_s(K) \left[
\ln \left( \frac{\Lambda^2}{-k^2} \right) + {\rm finite} \right]
+ {\cal O}(\alpha_s^2) \ .
\label{brackets}
\end{equation}
Thus, using the explicit form (\ref{kA}) for $\kappa_A$,
expression (\ref{bDS21}) can be rewritten as
\begin{eqnarray}
&&- k^2 ~c^a(k)^{(2)} + i~f^{abc} ~k^{\mu}
~\int~\frac{d^{4} k_1}{(2 \pi)^4} A^{b}_{\mu}(k_1)^{(1)}
c^c(k-k_1)^{(1)} =
\nonumber\\
&& \!\! \!\! \!\!\!\! \!\! 
- \frac{i}{4} (3 + \xi ^{(\Lambda)}) ~N_c ~\alpha_s^2(-k^2)~k_{\mu}
~\int^{(\Lambda)}~\frac{d^{4} k_1}{(2 \pi)^4}
\frac{1}{k_1^2 (k - k_1)^2} \left[
\ln \left( \frac{\Lambda^2}{-k^2} \right) + {\rm finite} \right]
\left( f^{abc} J^{\mu}_b(k_1) {\rho}^c(k-k_1) \right)
+ {\cal O}(\alpha_s^{5/2}) \ .
\label{bDS22}
\end{eqnarray}
The calculation of the last term in DSE (\ref{bDS2}),
i.e., its reduction to a form of the type (\ref{bDS22}),
is more involved since it entails evaluation of
(a variational derivative of) 
the cubic-in-sources function $c^{(3)}$. The
calculation is given in Appendix \ref{app:dc3dJ},
in the Feynman gauge ($\xi^{(\Lambda)}\!=\!1$), with the
result given in Eq.~(\ref{dc35}). Since the sum in the
first brackets there is one, we see that the third term 
result (\ref{dc35}) is just the negative of the
result (\ref{bDS22}) for the first two terms in
the Feynman gauge.\footnote{
The ``finite'', i.e. $\Lambda$-independent
parts, in addition to $\ln ( \Lambda^2/(-k^2))$,
may also appear in Eqs.~(\ref{dc35}) and (\ref{bDS22}),
the algebra of including them becomes very involved,
and they are not considered in the present analysis.}
This explicitly shows that the $\Lambda^2$- and 
$k^2$-dependence of the one-loop dressing functions
(\ref{GX1lall}) is consistent with the
quadratic-in-sources antighost DSE (\ref{bDS2})
in our effective action framework.

\section{Summary}
\label{sec:summary}

Based on arguments presented in our previous
work Ref.~\cite{KCS}, we propose a specific
form (\ref{GQCD}) of the QCD effective action
which basically repeats the structure of the
classical QCD action, but where all the fields $X$
are replaced by their convolutions with dressing
functions $G_X$: 
${\tilde X} = G_X^{-1} \circ X$.
The proposed pure gluon part is given in
Eq.~(\ref{Ggl1}), and we argue that it should
be supplemented by terms which restore gauge
invariance in the dressed gluon field ${\tilde A}$
and that these supplementary pure gluon terms
do not contribute to perturbative effects.
In general, the proposed effective action
should have all perturbative and nonperturbative
effects (except for possible soliton-like vacuum effects)
encoded in the dressing functions $G_X$.

We then investigated the consistency
of the proposed framework in the regime of 
application of perturbative
QCD (pQCD). First we deduced the gluon and ghost
dressing functions $G_A$ and $G_c$ from the requirement
that the full gluon and antighost-ghost propagators
of the framework should agree with those known
from pQCD (with an UV regularization scale $\Lambda$),
and obtained explicit pQCD expressions for $G_A$ and $G_c$ at
the one-loop level.
Stated differently, we fixed the two dressing
functions of the framework by requiring the
agreement of two two-point Green functions
calculated in this framework with those of pQCD. 
But the two dressing functions appear also in many other
(three-point, four-point, ...) Green functions.
These Green functions can be related
with the two-point ones via Dyson--Schwinger
equations (DSE's). A priori, it is far from clear
whether a very limited number of dressing
functions fulfills many (infinite number of)
DSE's. We performed explicit checks of the
antighost DSE ($b$-DSE) at the linear- and at the
quadratic-in-sources level, with the
aforementioned one-loop dressing functions
$G_A$ and $G_c$. These two
DSE's involve implicitly the connected three-point
$\langle b c A \rangle$ and four-point $\langle b c A A \rangle$
Green functions.
Thus we presented nontrivial checks of
the consistency of the proposed QCD
framework in the perturbative (high momentum) region.

However, in the quadratic-in-sources
DSE we considered only the large logarithms
containing the regularization scale [$\ln(\Lambda^2/(-k^2)]$
in the dressing functions and in the DSE's,
leaving the algebraically involved problem of the cancelation of the
$\Lambda$-independent (``finite'') terms in the DSE's 
for future investigation. Subsequently, we plan
to investigate, in the proposed QCD effective action
framework, the behavior of the quark sector 
in the high momentum region,
and subsequently the behavior of all sectors
in the low momentum region.

\begin{acknowledgments}
The work of G.C. and I.S. was supported by FONDECYT (Chile) 
grant No. 1010094 and  8000017, respectively.
The work of I.K. was supported by the Programa MECESUP FSM9901 
of the Ministry of Education (Chile) and also by 
CONICYT (Chile) under grant 8000017. 
\end{acknowledgments}

\appendix

\section{Logarithm of the (covariant) d'alambertian}
\label{app:lndel}

In this Appendix we write down explicit expressions for the
logarithm of d'alambertian $\partial^2 \equiv 
\partial_{\mu} \partial^{\mu}$ and 
the logarithm of its (tilde-)covariant version 
$\nabla^2(\tilde A) \equiv
\nabla_{\mu}(\tilde A) \nabla^{\mu}(\tilde A)$,
and clarify why expression (\ref{Ggl11l}) is
not gauge invariant and expression (\ref{Ggl1l}) is.

For any function $B(x)$ in spacetime, the function
$[\ln(\partial^2/\mu_0^2) B](x)$ can be written in terms of
the Taylor expansion of the logarithm around the point
$\ln(\mu_0^2/\mu_0^2) = 1$
\begin{eqnarray}
&&\left[ \ln \left( \frac{ \partial^2}{\mu_0^2} \right) B \right](x) 
= \left[ \frac{1}{\mu_0^2} (\partial^2 - \mu_0^2) 
- \frac{1}{2 \mu_0^4} (\partial^2 - \mu_0^2)^2 
+ \frac{1}{3 \mu_0^6} (\partial^2 - \mu_0^2)^3 + \cdots \right] B(x) \ .
\label{lndel2Bx}
\end{eqnarray}
The Fourier transformation (\ref{FTGX}) 
of this function is obtained by
integration by parts which replaces $\partial^2 \mapsto - k^2$
\begin{eqnarray}
\left[ \ln \left( \frac{ \partial^2}{\mu_0^2} \right) B \right] (k)
& = & \left[ - \frac{1}{\mu_0^2}(k^2 + \mu_0^2) 
- \frac{1}{2 \mu_0^4} (k^2 + \mu_0^2)^2 
- \frac{1}{3 \mu_0^6} (k^2 + \mu_0^2)^3 + \cdots \right] B(k)
\nonumber\\
& = & \ln \left( \frac{ - k^2}{\mu_0^2} \right) B(k) \ .
\label{lndel2Bb}
\end{eqnarray}
This shows that the one-loop relation (\ref{1lRGEall}) 
inserted in expression (\ref{Ggl1}) gives expression
(\ref{Ggl11l}). Further, 
since $\partial^2 F^{\mu \nu}({\tilde A})(x)$
is not gauge covariant [$F^{\mu \nu}({\tilde A})(x)$ is
gauge covariant], expression
${\rm Tr} [ F_{\mu \nu}({\tilde A})(x)
\partial^2 F^{\mu \nu}({\tilde A})(x) ]$ is not gauge
invariant, and then Eq.~(\ref{lndel2Bx}) implies that
the second integrand on the right-hand side of
Eq.~(\ref{Ggl11l}) is not gauge invariant.

Unlike $\partial^2 F^{\mu \nu}({\tilde A})(x)$,
expression
$\nabla^2({\tilde A}) F^{\mu \nu}({\tilde A})(x)$
is gauge covariant.
Therefore,
${\rm Tr} [ F_{\mu \nu}({\tilde A})(x)
\nabla^2({\tilde A}) F^{\mu \nu}({\tilde A})(x) ]$
is gauge invariant. The logarithmic version
can be written in complete analogy with Eq.~(\ref{lndel2Bx})
\begin{eqnarray}
&&\left[ \ln \left( \frac{ \nabla^2({\tilde A})}{\mu_0^2} 
\right) B \right](x) 
= \left[ \frac{1}{\mu_0^2} (\nabla^2({\tilde A}) - \mu_0^2) 
- \frac{1}{2 \mu_0^4} (\nabla^2({\tilde A})^2 - \mu_0^2)^2 
+ \frac{1}{3 \mu_0^6} (\nabla^2({\tilde A}) - \mu_0^2)^3 
+ \cdots \right] B(x) \ .
\label{lnDel2Bx}
\end{eqnarray}
{}From here we see that, when $B(x) = F^{\mu \nu}({\tilde A})(x)$,
this expression is also gauge covariant, and thus the
trace representing the second integrand 
on the right-hand side of Eq.~(\ref{Ggl1l})
is gauge invariant.

{}From this procedure, we can also see how to proceed
in the general (nonperturbative) case, i.e., when
$1/g(-k^2) \equiv {\cal G}(-k^2)$ is a general function
of $k^2$, in order to obtain from (\ref{Ggl1}) the
(tilde-)gauge invariant version
\begin{subequations}
\label{Gglall}
\begin{eqnarray}
\Gamma_{\rm gl,1}[A] &=&  - \frac{1}{2}~
\int~\frac{d^{4}k}{(2 \pi)^4}~{\cal G}(-k^2)~
{\rm Tr} \left[ 
F_{\mu \nu}({\tilde A})(k)~F^{\mu \nu}({\tilde A})(-k) 
\right] \quad \left( {\cal G}(-k^2) \equiv \frac{1}{g(-k^2)} \right)
\label{Ggl1F}
\\
\Rightarrow \ \Gamma_{\rm gl}[A] &=& 
- \frac{1}{2}~\int~d^{4}x~{\rm Tr} 
\left[F_{\mu \nu}({\tilde A})(x)~
{\cal G}\left( \nabla^2({\tilde A}(x)) \right)~
F^{\mu \nu}({\tilde A})(x) \right] \ ,
\label{Ggl}
\end{eqnarray}
\end{subequations}
where ${\cal G}(-k^2)$ should be represented as a power
expansion.

\section{Regularized integrals}
\label{app:regint}

In this article, we need the regularized
form of the following types of integrals:
\begin{subequations}
\label{Idefall}
\begin{eqnarray}
I(q_1)^{ \{1; \mu_1 \}} & = & 
\int^{(\Lambda)}~\frac{d^{4} k_1}{(2 \pi)^4}~
\frac{1}{k_1^2 (k_1 + q_1)^2}~ \left\{ 1; k_1^{\mu_1} \right\} \ ,
\label{I1def}
\\
I(q_1, q_2)^{ \{1; \mu_1; \mu_1 \mu_2; \mu_1 \mu_2 \mu_3 \} } & = & 
\int^{(\Lambda)}~\frac{d^{4} k_1}{(2 \pi)^4}~
\frac{1}{k_1^2 (k_1 + q_1)^2 (k_1 + q_2)^2}~ \left\{ 
1; k_1^{\mu_1};  k_1^{\mu_1} k_1^{\mu_2};
k_1^{\mu_1}k_1^{\mu_2}k_1^{\mu_3} \right\} \ ,
\label{I2def}
\end{eqnarray}
\end{subequations}
where the various options are indicated in the curly brackets.
The factors in the denominators are understood to have
$+ i~\varepsilon$ added to ensure causality.
The UV regularization is denoted in (\ref{Idefall})
by $(\Lambda)$ above the integration symbol.
It is implemented by dimensional regularization, 
i.e., in $D$ dimensions,
thus replacing $d^{4} k_1/(2 \pi)^4 \mapsto d^{D} k_1/(2 \pi)^D$
and $g_{\mu \nu} g^{\mu \nu} = 4 \mapsto D$.
The UV cutoff $\Lambda$ is introduced in the final results,
after $D \to 4$, according to the substitution
\begin{equation}
\frac{2}{4 - D} - \gamma + \ln ( 4 \pi) = \ln (\Lambda^2) \ ,
\label{LbarMS}
\end{equation}
representing thus the effective ``bare'' UV cutoff in
${\overline {\rm MS}}$ scheme.
The relevant formulas leading to the evaluation of
integrals (\ref{Idefall}) can be found, for example,
in Refs.~\cite{Muta,PS}.

The product of denominators is transformed into
one denominator via the method of Feynman parameters
\begin{equation}
\frac{1}{A_1 A_2 \ldots A_n} = \int_0^1~dx_1 \ldots dx_n~
\delta \left( \sum_1^n x_j - 1 \right)
\frac{ (n-1)!}{ [ x_1 A_1 + \cdots x_n A_n]^n} \ .
\label{Feynpar}
\end{equation}
Specifically,
\begin{subequations}
\label{FP2all}
\begin{eqnarray}
\frac{1}{k_1^2 (k_1 + q_1)^2} & = & \int_0^1 \int_0^1 ~dx_1 dx_2~
\delta(x_1+x_2-1) \frac{1}{\left( \ell^2 - \Delta^{(2)} \right)^2} \ ,
\label{FP2a}
\\
\ell & = & k_1 + x_2 q_1 \ , \qquad
\Delta^{(2)} = x_2 (1 - x_2) (- q_1^2) \ ;
\label{FP2b}
\end{eqnarray}
\end{subequations}
and
\begin{subequations}
\label{FP3all}
\begin{eqnarray}
\frac{1}{k_1^2 (k_1 + q_1)^2 (k_1 + q_2)^2} & = & 
2~\int_0^1 \int_0^1 \int_0^1~dx_1 dx_2 dx_3~
\delta(x_1+x_2+ x_3-1) 
\frac{1}{\left( \ell^2 - \Delta^{(3)} \right)^3} \ ,
\label{FP3a}
\\
\ell & = & k_1 + x_2 q_1 + x_3 q_2 \ , \qquad
\Delta^{(3)} = - x_1 x_2 q_1^2 - x_1 x_3 q_2^2 - x_2 x_3 (q_1 - q_2)^2
 \ .
\label{FP3b}
\end{eqnarray}
\end{subequations}
Introducing the integration variable $\ell$ according to
(\ref{FP2b}), (\ref{FP3b}), the integrals (\ref{Idefall})
can be rewritten as
\begin{subequations}
\label{Irewall}
\begin{eqnarray}
I(q_1)^{\{1; \mu_1 \}}  &=&  \lim_{D \to 4}
\int_0^1 \int_0^1~dx_1 dx_2~
\delta(x_1+x_2-1)~\int~\frac{d^{D} \ell}{ (2 \pi)^D} 
\frac{1}{\left( \ell^2  - \Delta^{(2)} \right)^2} 
\nonumber\\
&& \times \left\{ 1; \ell^{\mu_1}\!-\!x_2 q_1^{\mu_1} \right\} \ ,
\label{I1rew}
\\
I(q_1, q_2)^{ \{ 1; \mu_1;\mu_1 \mu_2; \mu_1 \mu_2 \mu_3 \} } & = & 
2~\lim_{D \to 4}
\int_0^1 \int_0^1 \int_0^1~dx_1 dx_2 dx_3~
\delta(x_1+x_2+x_3-1)
\nonumber\\
&& \times
\int~\frac{d^{D} \ell}{ (2 \pi)^D} 
\frac{1}{\left( \ell^2  - \Delta^{(3)} \right)^3}
\left\{ 1; q^{\mu_1}; q^{\mu_1} q^{\mu_2}; 
q^{\mu_1} q^{\mu_2} q^{\mu_3} \right\} 
{\big|}_{q \equiv l - x_2 q_1 - x_3 q_2} 
\ .
\label{I2rew}
\end{eqnarray}
\end{subequations}
The integration over $\ell$ can be performed after
Wick-rotation, using dimensional regularization,
and introducing $\epsilon \equiv (4\!-\!D)/2$
\begin{subequations}
\label{intellall}
\begin{eqnarray}
\int~\frac{d^{D} \ell}{ (2 \pi)^D}
\frac{1}{\left( \ell^2  - \Delta \right)^2} 
&  = & \frac{i}{(4 \pi)^2} \left[ \left( \frac{1}{\epsilon} - \gamma
+ \ln(4 \pi) - \ln(\Delta) \right) + {\cal O}(\epsilon) \right] \ ,
\label{intell1}
\\
\int~\frac{d^{D} \ell}{ (2 \pi)^D}
\frac{\ell^2}{\left( \ell^2  - \Delta \right)^3} 
&  = & \frac{i}{(4 \pi)^2} \left( 1 - \frac{\epsilon}{2} \right)
\left[ \left( \frac{1}{\epsilon} - \gamma
+ \ln(4 \pi) - \ln(\Delta) \right) + {\cal O}(\epsilon) \right] \ ,
\label{intell2}
\\
\int~\frac{d^{D} \ell}{ (2 \pi)^D}
\frac{1}{\left( \ell^2  - \Delta \right)^3} 
&  = & 
- \frac{i}{(4 \pi)^2} \frac{1}{2} \frac{1}{\Delta} + {\cal O}(\epsilon)
\quad {\rm (finite)} \ .
\label{intell3}
\end{eqnarray}
\end{subequations}
Further, by symmetry considerations the
following identities hold:
\begin{subequations}
\label{symmrel}
\begin{eqnarray}
&&\int~\frac{d^{D} \ell}{ (2 \pi)^D} 
\frac{ \{ \ell^{\mu_1}; \ell^{\mu_1}\ell^{\mu_2}\ell^{\mu_3} \} }
{\left( \ell^2  - \Delta \right)^n} = 0 \ ,
\\
&&\int~\frac{d^{D} \ell}{ (2 \pi)^D} 
\frac{ \ell^{\mu_1}\ell^{\mu_2} }
{\left( \ell^2  - \Delta \right)^n} =
\frac{1}{D} g^{\mu_1 \mu_2} 
\int~\frac{d^{4} \ell}{ (2 \pi)^4}
\frac{\ell^2}{\left( \ell^2  - \Delta \right)^n} \ .
\end{eqnarray}
\end{subequations}
Applying the identities (\ref{intellall})--(\ref{symmrel})
to (\ref{Irewall}), and at the end performing the limit
$\epsilon \equiv (4\!-\!D)/2 \to 0$, and then introducing
the cutoff notation (\ref{LbarMS}), we obtain the 
regularized integrals (\ref{Idefall})
\begin{subequations}
\label{Iall}
\begin{eqnarray}
I(q) &=& \frac{i}{(4 \pi)^2} \left[ \ln \left(
\frac{\Lambda^2}{- q^2} \right) + 2 \right] \ , 
\label{I11}
\\
I(q)^{\mu} &=& - q^{\mu} \frac{i}{2 (4 \pi)^2}
\left[ \ln \left(
\frac{\Lambda^2}{- q^2} \right) + 2 \right] \ ,
\label{I12}
\\
I(q_1,q_2) & = & - \frac{i}{ (4 \pi)^2}~
\int_0^1 d x_3~\int_0^{1 - x_3} d x_2~\frac{1}{F(q_1,q_2;x_2,x_3)}
\ ,
\label{I21}
\\
I(q_1,q_2)^{\mu} & = & \frac{i}{ (4 \pi)^2}~
\int_0^1 d x_3~\int_0^{1 - x_3} d x_2
~\frac{( x_2 q_1^{\mu}\!+\!x_3 q_2^{\mu} )}{F(q_1,q_2;x_2,x_3)}
\ ,
\label{I22}
\\
I(q_1,q_2)^{ \mu_1 \mu_2} &=&
\frac{i}{ (4 \pi)^2} {\Bigg \{} \frac{1}{4} g^{\mu_1 \mu_2}
\ln \left( \frac{ \Lambda^2}{ - q^2} \right)
- \frac{1}{2} g^{\mu_1 \mu_2}~
\int_0^1 d x_3~\int_0^{1 - x_3} d x_2~\ln \left[
\frac{F(q_1,q_2;x_2,x_3)}{- q^2} \right] 
\nonumber\\
&&- \int_0^1 d x_3~\int_0^{1 - x_3} d x_2
~\frac{( x_2 q_1\!+\!x_3 q_2 )^{\mu_1}
( x_2 q_1\!+\!x_3 q_2 )^{\mu_2}}{F(q_1,q_2;x_2,x_3)} 
{\Bigg \}} \ ,
\label{I23}
\\
I(q_1,q_2)^{\mu_1 \mu_2 \mu_3} &=&
\frac{i}{ (4 \pi)^2} {\Bigg \{}
- \frac{1}{12} \left[ g^{\mu_1 \mu_2} (q_1\!+\!q_2)^{\mu_3} +
g^{\mu_2 \mu_3} (q_1\!+\!q_2)^{\mu_1} +
g^{\mu_3 \mu_1} (q_1\!+\!q_2)^{\mu_2} \right]
~\ln \left( \frac{ \Lambda^2}{ - q^2} \right) 
\nonumber\\
&& 
\!\!\!\!\!\!\!\!\!\!\!\!\!\!\!\!\!\!\!\!\!\!\!\!\!\!\!\!\!\!\!\!\!\!\!\!
\!\!\!\!\!
+ \frac{1}{2} ~\int_0^1 d x_3~\int_0^{1 - x_3} d x_2 
\left[g^{\mu_1 \mu_2} (x_2 q_1\!+\!x_3 q_2)^{\mu_3} +
g^{\mu_2 \mu_3} (x_2 q_1\!+\!x_3 q_2)^{\mu_1} +
g^{\mu_3 \mu_1} (x_2 q_1\!+\!x_3 q_2)^{\mu_2} \right]
~\ln \left[ \frac{ F(q_1,q_2;x_2,x_3)}{- q^2} \right]
\nonumber\\
&& 
\!\!\!\!\!\!\!\!\!\!\!\!
+  ~\int_0^1 d x_3~\int_0^{1 - x_3} d x_2
~\frac{( x_2 q_1\!+\!x_3 q_2 )^{\mu_1}
( x_2 q_1\!+\!x_3 q_2 )^{\mu_2} ( x_2 q_1\!+\!x_3 q_2 )^{\mu_3} }
{F(q_1,q_2;x_2,x_3)}  {\Bigg \}} \ ,
\label{I24}
\end{eqnarray}
\end{subequations}
where we denoted
\begin{equation}
F(q_1,q_2;x_2,x_3) \equiv - x_2 (1\!-\!x_2) q_1^2
- x_3 (1\!-\!x_3) q_2^2 + 2 x_2 x_3 (q_1 \cdot q_2) \ ,
\label{Fdenote}
\end{equation}
and the squared scale $(-q^2)$ appearing in
Eqs.~(\ref{I23})--(\ref{I24}) is arbitrary, e.g.,
$-q^2 = -q_1^2$ or $-q^2 = -q_2^2$ [the results
(\ref{I23})--(\ref{I24}) are independent of $(- q^2)$ ].
In the simpler cases when only one fixed scale appears in
the integrals, say $q_1=0$ and $q_2=q$, we obtain from the
above results immediately
\begin{subequations}
\label{Iball}
\begin{eqnarray}
I(0,q) & = & \frac{i}{ (4 \pi)^2} \frac{1}{q^2}~\int_{\eta}^1
\frac{d x_3}{x_3} \quad (\eta \to +0) \ ,
\label{I21b}
\\
I(0,q)^{\mu} & = & - \frac{i}{ (4 \pi)^2} \frac{q^{\mu}}{q^2} \ ,
\label{I22b}
\\
I(0,q)^{\mu_1 \mu_2} & = & \frac{i}{ 2 (4 \pi)^2}
\left[ \frac{1}{2} 
g^{\mu_1 \mu_2} \ln \left( \frac{ \Lambda^2}{-q^2} \right)
+ g^{\mu_1 \mu_2} + \frac{ q^{\mu_1} q^{\mu_2}}{q^2} \right] \ ,
\label{I23b}
\\
I(0,q)^{\mu_1 \mu_2 \mu_3} & = & - \frac{i}{ 3 (4 \pi)^2}
{\Bigg \{} \frac{1}{4} \left( g^{\mu_1 \mu_2} q^{\mu_3} +
g^{\mu_2 \mu_3} q^{\mu_1} + g^{\mu_3 \mu_1} q^{\mu_2} \right)
\left[ \ln \left( \frac{ \Lambda^2}{-q^2} \right) + \frac{5}{3}
\right] + \frac{ q^{\mu_1} q^{\mu_2} q^{\mu_3} }{q^2}
{\Bigg \}} \ .
\label{I24b}
\end{eqnarray} 
\end{subequations}

As an application, using the results (\ref{Iall}), (\ref{Iball}),
we obtain the integrals on the
right-hand side of DSE (\ref{bDS13})
\begin{eqnarray}
&& \int^{(\Lambda)}~\frac{d^{4} k_1}{(2 \pi)^4}
~\frac{1}{k_1^2 (k-k_1)^2} k^{\mu_1} g_{\mu_1 \mu_2} k_1^{\mu_2}
= I(-k)^{\mu_2} k_{\mu_2} = \frac{i}{2 (4 \pi)^2}~k^2~
\left[ \ln \left( \frac{\Lambda^2}{- k^2} \right) + 2 \right]
\ ,
\label{bDSEint11}
\\
&& \int^{(\Lambda)}~\frac{d^{4} k_1}{(2 \pi)^4}
~\frac{k^{\mu_1}}{k_1^2 (k-k_1)^2~}  
\frac{(k\!-\!k_1)_{\mu_1} (k\!-\!k_1)_{\mu_2}}{(k-k_1)^2}
~k_1^{\mu_2}  = \int^{(\Lambda)}~\frac{d^{4} q_1}{(2 \pi)^4}
~\frac{k^{\mu_1} q_{\mu_1} q_{\mu_2} (k + q_1)^{\mu_2}}
{q_1^2 q_1^2 (q_1 + k)^2}
\nonumber\\
&& = k^{\mu_1} k^{\mu_2} I(0,k)_{\mu_1 \mu_2} +
k^{\mu_1} I(0,k)_{\mu_1 \mu_2}^{\;\;\;\;\;\;\;\;\mu_2}
\nonumber\\
&& = \frac{i}{ (4 \pi)^2} k^2 {\Bigg \{}
\left[ \frac{1}{4} \ln \left( \frac{ \Lambda^2}{-k^2} \right) + 1
\right] - \frac{1}{3} \left[ \frac{1}{4} \left( D - 4 + 6 \right)
\left( \ln \left( \frac{ \Lambda^2}{-k^2} \right) + \frac{5}{3}
\right) + 1 \right] {\Bigg\}}
\label{bDSEintint12}
\\
& = & - \frac{i}{ 4 (4 \pi)^2} k^2 
\ln \left( \frac{ \Lambda^2}{-k^2} \right) \ .
\label{bDSEint12}
\end{eqnarray}
The step from Eq.~(\ref{bDSEintint12}) to Eq.~(\ref{bDSEint12})
follows by using the identity (\ref{LbarMS}).

Combining (\ref{bDSEint11}) and (\ref{bDSEint12}),
we obtain the result for the regularized expression on the 
right-hand side of the DSE (\ref{bDS13})
\begin{eqnarray}
&& i~4 \pi \alpha_s(K) N_c \frac{1}{k^2} 
~\int^{(\Lambda)}~\frac{d^{4} k_1}{(2 \pi)^4}
~\frac{1}{k_1^2 (k-k_1)^2}
k^{\mu_1} 
\left[ g_{\mu_1 \mu_2} + (\xi^{(\Lambda)}\!-\!1)
\frac{(k\!-\!k_1)_{\mu_1} (k\!-\!k_1)_{\mu_2}}{(k-k_1)^2}
\right] k_1^{\mu_2}  = 
\nonumber\\
&& 
\;\;\;\;\;\;\;\;\;\;\;\;\;
- \alpha_s(K) ~N_c ~\frac{1}{16 \pi}
~\left[ (3 - \xi^{(\Lambda)})
\ln \left( \frac{\Lambda^2}{- k^2} \right) 
+ 4  \right] \ .
\label{bDS13rhs}
\end{eqnarray}

\section{Evaluation of the variational derivative of $c^{(3)}$}
\label{app:dc3dJ}

In this Appendix, we evaluate, i.e., reduce to the form of
the type (\ref{bDS22}), the last term in the
quadratic-in-sources antighost DSE (\ref{bDS2}).
This term can be first rewritten as an integral over two
momenta, by using the expression (\ref{c3})
\begin{eqnarray}
\lefteqn{
f^{a c_2 c_3} k^{\mu_1} \left( 
\frac{\delta c^{c_3 (3)}}{\delta J^{\mu_1}_{c_2}} \right) (k) =
i f^{a c_2 c_3} f^{c_3 b_2 b_3} k^{\mu_1}~\int^{(\Lambda)}
~\frac{d^{4} k_1 d^{4}k_2}{ (2 \pi)^4} \frac{1}{(k_1+k)^2} 
\frac{ G_c((k_1+k)^2)}{G_A(k_2^2) G_c( (k_1 + k - k_2)^2)} 
(k_1 + k)^{\mu_2}
}
\nonumber\\
&& \times
\left[ \frac{ \delta A^{b_2}_{\mu_2}(k_2)^{(1)}}
{\delta J^{\mu_1}_{c_2}(k_1)} c^{b_3}(k_1 + k - k_2)^{(2)} +  
A^{b_2}_{\mu_2}(k_2)^{(1)}
\frac{ \delta c^{b_3}(k_1 + k - k_2)^{(2)}}
{\delta J^{\mu_1}_{c_2}(k_1)} 
+ \frac{ \delta A^{b_2}_{\mu_2}(k_2)^{(2)}}
{\delta J^{\mu_1}_{c_2}(k_1)} c^{b_3}(k_1 + k - k_2)^{(1)}
\right] \ .
\label{dc31}
\end{eqnarray}
We note that $(\delta c^{(3)}/\delta J)(k)$
here means the Fourier transform of the $x$-function
$(\delta c^{(3)}(x)/\delta J(x))$.
We then insert for the three terms in integral (\ref{dc31}) the
expressions for $c^{(1)}, c^{(2)}, A^{(1)}, A^{(2)}$
in terms of the dressing functions $G_X$ and the
sources, as given in relations (\ref{c1})--(\ref{c2})
and (\ref{A1})--(\ref{A2}). This results in the corresponding
three integrals over two momenta, quadratic in the
sources ($\sim J {\rho}$), and containing the dressing
functions $G_A$ and $G_c$ and the gluon propagators 
$P_{\mu_i \mu_j}$ (\ref{Pdef})
\begin{eqnarray}
\lefteqn{
f^{a c_2 c_3} k^{\mu_1} \left( 
\frac{\delta c^{c_3 (3)}}{\delta J^{\mu_1}_{c_2}} \right) (k) =
}
\nonumber\\
&&
- g^4(K) f^{a c_2 c_3} f^{c_3 c_2 b_3} f^{b_3 d_2 d_3} \frac{1}{k^2}
~\int^{(\Lambda)}~\frac{d^{4} k_1 d^{4}k_2}{ (2 \pi)^8}
\frac{G_A(k_1^2) G_c((k_1+k)^2) G_A(k_2^2)}{
k_1^2 (k_1+k)^2 k_2^2 (k_2-k)^2 G_A((k_2-k)^2)}
\nonumber\\
&& \times
k^{\mu_1} (k_1+k)^{\mu_2} P_{\mu_2 \mu_1}(k_1) k^{\mu_3}
P_{\mu_3 \mu_4}(k_2) J_{d_2}^{\mu_4}(k_2) {\rho}^{d_3}(k-k_2)
\nonumber\\
&&
- g^4(K) f^{a c_2 c_3} f^{c_3 b_2 b_3} f^{b_3 c_2 d_3}
~\int^{(\Lambda)}~\frac{d^{4} k_1 d^{4}k_2}{ (2 \pi)^8}
\frac{G_A(k_1^2) G_c((k_1+k)^2) G_A(k_2^2)}{
k_1^2 (k_1+k)^2 k_2^2 (k_2-k)^2 (k_1 + k - k_2)^2 G_A((k_2-k)^2)}
\nonumber\\
&& \times
k^{\mu_1} (k_1+k)^{\mu_2} P_{\mu_2 \mu_4}(k_2) (k_1 + k - k_2)^{\mu_3}
P_{\mu_3 \mu_1}(k_1) J_{b_2}^{\mu_4}(k_2) {\rho}^{d_3}(k-k_2)
\nonumber\\
&&
- g^4(K) f^{a c_2 c_3} f^{c_3 b_2 b_3} f^{b_2 c_2 c_3}
~\int^{(\Lambda)}~\frac{d^{4} k_1 d^{4}k_2}{ (2 \pi)^8}
\frac{G_A(k_1^2) G_c((k_1+k)^2) G_A((k_2-k_1)^2)}{
k_1^2 (k_1+k)^2 k_2^2 (k_2-k_1)^2 (k_1 + k - k_2)^2 
G_A((k_1 + k -k_2)^2)}
\nonumber\\
&& \times
k^{\mu_1}(k_1+k)^{\mu_2} P_{\mu_2 \mu_3}(k_2)
\bigg[ (2 k_1 - k_2)^{\mu_3} P_{\mu_4 \mu_1}(k_1)
P^{\mu_4}_{\;\;\;\mu_5}(k_2 - k_1) +
(2 k_2 - k_1)^{\mu_4} P_{\mu_4 \mu_1}(k_1) 
P^{\mu_3}_{\;\;\;\mu_5}(k_2 - k_1)
\nonumber\\
&& - (k_1 + k_2)^{\mu_4} 
P^{\mu_3}_{\;\;\;\mu_1}(k_1) P_{\mu_4 \mu_5}(k_2 - k_1)
\bigg] J_{c_3}^{\mu_5}(k_2-k_1) {\rho}^{b_3}(k_1 + k -k_2) \ .
\label{dc32}
\end{eqnarray}
The cubic terms in the structure constants can be reduced
due to the following identities:
\begin{eqnarray}
f^{a c_2 c_3} f^{c_3 c_2 b_3} &=& - N_c \delta_{a b_3} \ ,
\label{f1}
\\
f^{a c_2 c_3} f^{c_3 b_2 b_3} f^{b_3 c_2 d_3} &=& 
- \frac{1}{2} N_c f^{a b_2 d_3} \ , \quad
f^{a c_2 c_3} f^{c_3 b_2 b_3} f^{b_2 c_2 c_3} =
- \frac{1}{2} N_c f^{a c_3 b_3} \ .
\label{f2}
\end{eqnarray}
Identity (\ref{f1}) is well known, and
identities (\ref{f2}) are direct consequences of the
Jacobi identity. We use identities (\ref{f1}) and (\ref{f2}) 
in expressions (\ref{dc32}); further, we rename in
the second double integral $b_2 \mapsto d_2$ and $\mu_2
\leftrightarrow \mu_3$; in the third double integral,
$c_3 \mapsto d_2$, $b_3 \mapsto d_3$,
$\mu_4 \leftrightarrow \mu_5$ and $k_2({\rm new}) = k_2 - k_1$.
This allows us to write expressions (\ref{dc32})
in a somewhat more compact form
\begin{eqnarray}
\lefteqn{
f^{a c_2 c_3} k^{\mu_1} \left( 
\frac{\delta c^{c_3 (3)}}{\delta J^{\mu_1}_{c_2}} \right) (k) =
}
\nonumber\\
&& g^4(K) N_c ~\int^{(\Lambda)}~\frac{d^{4} k_1 d^{4}k_2}{ (2 \pi)^8}
~\frac{G_A(k_1^2) G_c((k_1+k)^2) G_A(k_2^2)}{G_A((k_2-k)^2)}
~\Bigg\{ \frac{1}{k^2} 
\frac{k^{\mu_1} P_{\mu_1 \mu_2}(k_1) (k_1+k)^{\mu_2} k^{\mu_3}
P_{\mu_3 \mu_4}(k_2)}{k_1^2 (k_1+k)^2 k_2^2 (k_2-k)^2}
\nonumber\\
&&
+ \frac{1}{2} \frac{k^{\mu_1} 
P_{\mu_1 \mu_2}(k_1) (k_1 + k - k_2)^{\mu_2}
(k_1+k)^{\mu_3} P_{\mu_3 \mu_4}(k_2)} 
{k_1^2 (k_1+k)^2 k_2^2 (k_2-k)^2 (k_1 + k - k_2)^2}
\nonumber\\
&&
+ \frac{1}{2} \frac{k^{\mu_1}(k_1+k)^{\mu_2} P_{\mu_2 \mu_3}(k_1+k_2)}
{k_1^2 (k_1+k)^2 k_2^2 (k_2+k_1)^2 (k_2 - k)^2}
\bigg[ (k_1 - k_2)^{\mu_3} P_{\mu_1 \mu_5}(k_1)
P^{\mu_5}_{\;\;\;\mu_4}(k_2) 
\nonumber\\
&&+ (2 k_2 + k_1)^{\mu_5} P_{\mu_5 \mu_1}(k_1) 
P^{\mu_3}_{\;\;\;\mu_4}(k_2)
- (2k_1 + k_2)^{\mu_5} P^{\mu_3}_{\;\;\;\mu_1}(k_1) 
P_{\mu_5 \mu_4}(k_2)
\bigg]
\Bigg\}
\times \left( f^{a d_2 d_3} J^{\mu_4}_{d_2}(k_2) {\rho}^{d_3}(k-k_2)
\right) \ .
\label{dc33}
\end{eqnarray}
The next step is to perform in Eq.~(\ref{dc33})
the integration over $k_1$.
Since we work in the one-loop approximation, we need to
obtain only the leading order term of expression (\ref{dc33}),
i.e., the terms $\sim g^3 \sim \alpha_s^{3/2}$ 
(note that $J \sim g^{-1}$),
because the one-loop expressions for the dressing functions
predict only the term $\sim \alpha_s^{3/2}$ in Eq.~(\ref{bDS22})
for the first two terms of the quadratic-in-sources antighost
DSE (\ref{bDS2}). Therefore, we replace in Eq.~(\ref{dc33})
all the dressing functions $G_X(k_s^2) \mapsto 1$. 
Nonetheless, the integration over 
$k_1$ is still involved, because of the rather
complicated momentum dependence of the gluon propagators
$P_{\mu_i \mu_j}(k_s)$, cf.~Eq.~(\ref{Pdef}). However, 
in the Feynman gauge, $\xi^{(\Lambda)} = 1$, and therefore
$\xi^{(K)} = 1 + {\cal O}(\alpha_s)$ by relation (\ref{xirun}).
So we can set $\xi^{(K)} \mapsto 1$ in the Feynman gauge
propagators in Eq.~(\ref{dc33}) at the considered order,
i.e., $P_{\mu_i \mu_j}(k_s) \mapsto g_{\mu_i \mu_j}$.
Thus, we obtain
\begin{eqnarray}
\lefteqn{
\left[ f^{a c_2 c_3} k^{\mu_1} \left( 
\frac{\delta c^{c_3 (3)}}{\delta J^{\mu_1}_{c_2}} \right) (k)
\right]_{\xi^{(\Lambda)} = 1} =
(16 \pi^2)~\alpha_s^2(K)~N_c~
\int^{(\Lambda)}~\frac{d^{4} k_1 d^{4}k_2}{ (2 \pi)^8}
}
\nonumber\\
&& \times
\Bigg\{
\frac{1}{k^2} \frac{\left(k \cdot (k_1 + k) \right) k_{\mu} }
{k_1^2 (k_1+k)^2 k_2^2 (k_2-k)^2}
+ \frac{1}{2} \frac{ \left( k \cdot (k_1 + k - k_2) \right) 
(k_1 + k)_{\mu} }{k_1^2 (k_1+k)^2 k_2^2 (k_2-k)^2 (k_1 + k - k_2)^2}
\nonumber\\
&& + \frac{1}{2} \frac{ \left[
\left( (k_1+k)\cdot (k_1 - k_2) \right) k_{\mu} +
\left( k \cdot (k_1 + 2 k_2) \right) (k_1 + k)_{\mu}
- \left( k \cdot (k_1 + k) \right) (2 k_1 + k_2)_{\mu}
\right] }{k_1^2 (k_1+k)^2 k_2^2 (k_2+k_1)^2 (k_2 - k)^2}
\Bigg\}
\nonumber\\
&&\times \left( f^{a d_2 d_3} J^{\mu}_{d_2}(k_2) {\rho}^{d_3}(k-k_2)
\right)  + {\cal O}(\alpha_s^{5/2}) \ .
\label{dc34}
\end{eqnarray}
We should first perform the integration over $k_1$, which is
complicated. Here we only take into account the cutoff-dependent
terms [$\sim\!\ln(\Lambda^2/(-q^2))$]. Then the integration
of each term over $k_1$ becomes less difficult, using the
results (\ref{I11})--(\ref{I23}) 
[the terms $\sim\!\ln(\Lambda^2/(-q^2))$ there ]. The integrals
needed for the first, second, and third term of Eq.~(\ref{dc34})
are respectively
\begin{subequations}
\label{fst}
\begin{eqnarray}
&&\int^{(\Lambda)}~\frac{d^{4} k_1}{ (2 \pi)^4}
\frac{ (k^2 + k_1^{\mu_1} k_{\mu_1}) k_{\mu} }{ 
k_1^2 (k_1 + k)^2} = 
I(k) k^2 k_{\mu} + 
I(k)^{\mu_1} k_{\mu_1} k_{\mu} =
k^2 k_{\mu} \frac{1}{2} \frac{i}{(4 \pi)^2}
\left[ \ln \left( \frac{ \Lambda^2}{-k^2} \right) + {\rm finite} 
\right] \ ,
\label{first}
\\
&&\int^{(\Lambda)}~\frac{d^{4} k_1}{ (2 \pi)^4}
\frac{ \left( k_{\mu_1} k_1^{\mu_1} + k \cdot (k-k_2)
\right) ( k_{1 \mu} + k_{\mu})  } 
{k_1^2 (k_1 + k)^2 (k_1 + k - k_2)^2} 
\nonumber\\
&& = I(k,k-k_2)^{\mu_1}_{\;\;\;\mu} k_{\mu_1} + {\rm finite}
= k_{\mu} \frac{1}{4} \frac{i}{(4 \pi)^2} \left[
\ln \left( \frac{ \Lambda^2}{-k^2} \right) + {\rm finite} 
\right] \ ,
\label{second}
\\
&&\int^{(\Lambda)}~\frac{d^{4} k_1}{ (2 \pi)^4}
\frac{ \left( k_1^{\mu_1} k_{1 \mu_1} k_{\mu} +
k_{\mu_1} k_1^{\mu_1} k_{1 \mu} - 2 k_{\mu_1} k_1^{\mu_1} k_{1 \mu}
\right) }{ k_1^2 (k_1 + k)^2 (k_1 + k_2)^2 } +
{\rm finite} 
\nonumber\\
&& =
I(k,k_2)^{\mu_1 \mu_2} g_{\mu_1 \mu_2} k_{\mu} -
I(k,k_2)^{\mu_1}_{\;\;\;\mu} k_{\mu_1} + {\rm finite}
 = k^{\mu} \frac{3}{4} \frac{i}{(4 \pi)^2} \left[
\ln \left( \frac{ \Lambda^2}{-k^2} \right) + {\rm finite} 
\right] \ .
\label{third}
\end{eqnarray}
\end{subequations}
Using the results of $k_1$-integration 
(\ref{fst}), we can rewrite Eq.~(\ref{dc34}) as
\begin{eqnarray}
\lefteqn{
\left[ f^{a c_2 c_3} k^{\mu_1} \left( 
\frac{\delta c^{c_3 (3)}}{\delta J^{\mu_1}_{c_2}} \right) (k)
\right]_{\xi^{(\Lambda)} = 1} =
(16 \pi^2)~\alpha_s^2(K)~N_c~
\int^{(\Lambda)}~\frac{d^{4}k_2}{ (2 \pi)^4}
\frac{1}{k_2^2 (k-k_2)^2} k_{\mu} \frac{i}{(4 \pi)^2}
}
\nonumber\\
&& \times
\left[ \frac{1}{k^2}\frac{1}{2} k^2 + \frac{1}{2} \frac{1}{4}
+ \frac{1}{2} \frac{3}{4} \right]
\left[ \ln \left( \frac{\Lambda^2}{-k^2} \right) + 
{\rm finite} \right]
\left( f^{a d_2 d_3} J^{\mu}_{d_2}(k_2) {\rho}^{d_3}(k-k_2)
\right)  + {\cal O}(\alpha_s^{5/2}) \ .
\label{dc35}
\end{eqnarray}
The sum in the first brackets is one.
This is the result for the third term of the
quadratic-in-sources antighost DSE (\ref{bDS2}),
in the Feynman gauge and at the leading
($\sim \alpha_s^{3/2}$) level.


\begin{thebibliography}{99}

\bibitem{ST}
A.~A.~Slavnov,
Theor.\ Math.\ Phys.\  {\bf 10}, 99 (1972)
[Teor.\ Mat.\ Fiz.\  {\bf 10}, 153 (1972)];
J.~C.~Taylor,
Nucl.\ Phys.\ B {\bf 33}, 436 (1971);
A.~A.~Slavnov,
Nucl.\ Phys.\ B {\bf 97}, 155 (1975).

\bibitem{SF}
A.~Slavnov and L.~Faddeev, Introduction to quantum theory
             of gauge fields, Moscow, Nauka, 1988.

\bibitem{Lee}
B.~W.~Lee,
Phys.\ Lett.\ B {\bf 46}, 214 (1973);
B.~W.~Lee,
Phys.\ Rev.\ D {\bf 9}, 933 (1974);
J.~Zinn--Justin, Lecture Note in Physics, v.37, Berlin,
Springer Verlag, 1974.

\bibitem{BRST}
C.~Becchi, A.~Rouet and R.~Stora,
Commun.\ Math.\ Phys.\  {\bf 42}, 127 (1975);
I.V. Tyutin, preprint FIAN, Gauge invariance in field theory
               and statistical physics in operator formalism, 
               Lebedev-75-39 (in Russian), 1975.

\bibitem{Becchi}
C.~Becchi,
Introduction to BRS symmetry,
arXiv:hep-th/9607181.

\bibitem{Kondrashuk:2000br}
I.~Kondrashuk,
JHEP {\bf 0011}, 034 (2000)
[arXiv:hep-th/0007136].

\bibitem{KCS}
I.~Kondrashuk, G.~Cveti\v{c}, and I.~Schmidt,
arXiv:hep-ph/0203014.

\bibitem{Muta}
T.~Muta, Foundations of quantum chromodynamics
(World Scientific, 1987).

\bibitem{PS}
M.~E.~Peskin and D.~V.~Schroeder, An introduction to
quantum field theory (Addison--Wesley,1996).

\end{thebibliography}
\end{document}